\documentclass[twocolumn,showkeys,amsmath,amssymb,pra]{revtex4}

\usepackage{graphicx,amssymb}
\usepackage{dcolumn}% Align table columns on decimal point
\usepackage{bm}% bold math

\newcommand{\beq}{\begin{equation}}
\newcommand{\eeq}{\end{equation}}
\newcommand{\bea}{\begin{eqnarray}}
\newcommand{\eea}{\end{eqnarray}}

\begin{document}

\title{Dynamical Hartree-Fock-Bogoliubov Theory of Vortices in Bose-Einstein
Condensates at Finite Temperature}

\author{B. G. Wild and D. A. W. Hutchinson}
\affiliation{The Jack Dodd Centre for Quantum Technology, Department of Physics, University of Otago, Dunedin 9054, New Zealand}

\begin{abstract}
We present a method utilizing the continuity equation for the condensate
density  to make predictions of the precessional frequency of single
off-axis vortices and of vortex arrays in Bose-Einstein condensates
at finite temperature. We also present an orthogonalized Hartree-Fock-Bogoliubov
(HFB) formalism. We solve the continuity equation for the condensate
density self-consistently with the orthogonalized HFB equations, and
find stationary solutions in the frame rotating at this frequency.
As an example of the utility of this formalism we obtain time-independent
solutions for quasi-two-dimensional rotating systems in the co-rotating
frame. We compare these results with time-dependent predictions where
we simulate stirring of the condensate.
\end{abstract}

\maketitle

\section{Introduction}

One of the characteristics of a superfluid is the quantization of
vortices that are found when sufficient angular momentum is present
in the system. Vortices, as a signature of superfluidity, are therefore
of great theoretical and experimental interest, and the reader is
referred to the recent review article by Fetter \cite{key-1}. Our
model is based on the Hartree-Fock-Bogoliubov (HFB) formalism \cite{key-2,key-3,key-4,key-5}.
We demonstrate how the continuity equation for the condensate density,
solved self-consistently with a set of time-independent orthogonalized
Hartree-Fock-Bogoliubov (HFB) equations in the frame rotating at the
precessional frequency, can be used to make \emph{a priori} predictions
of the precessional frequencies of vortices in Bose-Einstein condensates
(BECs). By introducing a set of modified basis functions incorporating
the vortex positions one is able to do this not only for single off-axis
vortices, but also for the multiple vortex case. In order to perform
these calculations correctly, it is necessary that the condensate
and thermal populations be mutually orthogonal. We present an orthogonal
HFB formalism in which this condition holds, and for which we show
the existence of a zero-energy eigenvalue in the time-independent
case, in contrast with the standard HFB formalism. As an illustration
we model a two-dimensional BEC system, establishing the dependency
of the precessional frequency $\Omega$ on the lattice parameter $a$
and on the temperature $T$ for triangular and hexagonal vortex arrays.
We also show the relationship of $\Omega$ with $T$ for the case
of two vortices and for triangular and hexagonal vortex arrays with
three and seven vortices, respectively, having lattice parameter $a=3$
harmonic oscillator units, establishing the existence of an upper
bound for the precessional frequency as a function of the number of
vortices. We obtain qualitative agreement with the areal density approximation
\cite{key-6,key-6a} (see also Ref. \cite{key-7}) for the hexagonal
vortex array. We verify the validity of these predictions in a series
of finite temperature simulations using time-dependent HFB \cite{key-8,key-9}.
We use one or more Gaussian stirrers to impart angular momentum to
the BEC, and establish a critical stirring frequency required for
the creation of vortices corresponding to a local stirring velocity
just in excess of the local Landau critical velocity. We also verify
that the axial component of the angular momentum is conserved when
the trapping potential is axially symmetric, thus satisfying Noether's
theorem. We show that angular momentum is lost when this symmetry
is broken, leading to the decay of vortices.

\section{Formalism}

We consider a BEC system in the frame rotating with angular frequency
$\mathbf{\Omega}$ with grand-canonical Bose Hamiltonian  given by\begin{equation}
\begin{array}{ccc}
\hat{H}^{(GC)}(t) & = & \int d\mathbf{r}\left[\hat{\psi}^{\dag}(\mathbf{r},t)\left(\hat{h}_{\Omega}(\mathbf{r})-\mu\right)\hat{\psi}(\mathbf{r},t)\right.\\
 &  & \left.+\frac{g}{2}\hat{\psi}^{\dag}(\mathbf{r},t)\hat{\psi}^{\dag}(\mathbf{r},t)\hat{\psi}(\mathbf{r},t)\hat{\psi}(\mathbf{r},t)\right].\end{array}\label{eq:1}\end{equation}
where $\hat{h}_{\Omega}(\mathbf{r})$ is the single-particle Hamiltonian\begin{equation}
\hat{h}_{\Omega}(\mathbf{r})=-\frac{\hbar^{2}}{2m}\nabla^{^{2}}+i\hbar\mathbf{\Omega}\cdot(\mathbf{r}\times\mathbf{\nabla})+V_{T}(\mathbf{r}),\label{eq:2}\end{equation}
and\begin{equation}
g=\frac{4\pi\hbar^{2}a_{s}}{m},\label{eq:3}\end{equation}
with $a_{s}$ the $s$-wave scattering length. We wish to obtain an
HFB formalism such that the condensate and thermal modes are orthogonal.
We proceed by splitting the Bose field operator $\hat{\psi}(\mathbf{r},t)$
into a coherent part represented by the condensate field operator
$\hat{\Phi}(\mathbf{r},t)$, and an incoherent part represented by
the fluctuation operator $\hat{\eta}(\mathbf{r},t)$, writing $\hat{\psi}(\mathbf{r},t)=\hat{\Phi}(\mathbf{r},t)+\hat{\eta}(\mathbf{r},t)$
with $\hat{\Phi}(\mathbf{r},t)\equiv\phi(\mathbf{r},t)\hat{a_{c}}(t)$,
with $\hat{a_{c}}(t)$ the annihilation operator for the condensate,
and $\phi(\mathbf{r},t)$ a condensate wave function satisfying the
normalization condition $\int d\mathbf{r}\left|\phi(\mathbf{r},t)\right|^{2}=1.$
The fluctuation operator $\hat{\eta}(\mathbf{r},t)$ is defined as
\cite{key-10} $\hat{\eta}(\mathbf{r},t)\equiv\hat{\psi}(\mathbf{r},t)-\phi(\mathbf{r},t)\int d\mathbf{r}\phi^{*}(\mathbf{r},t)\hat{\psi}(\mathbf{r},t)$.
It can then be shown that the orthogonality condition\begin{equation}
\int d\mathbf{r}\phi^{*}(\mathbf{r},t)\hat{\eta}(\mathbf{r},t)=0\label{eq:5}\end{equation}
 holds, so the condensate and thermal populations are orthogonal,
as required. We also find from the definition of $\hat{\eta}$ that
$\hat{a_{c}}(t)=\int d\mathbf{r}\phi^{*}(\mathbf{r},t)\hat{\psi}(\mathbf{r},t).$
In addition $\hat{\Phi}$ and $\hat{\eta}$ satisfy the commutation
relations\begin{equation}
\begin{array}{l}
\left[\hat{\Phi}(\mathbf{r},t),\hat{\psi}^{\dag}(\mathbf{r}^{\prime},t)\right]=\phi(\mathbf{r},t)\phi^{*}(\mathbf{r}^{\prime},t),\\
\left[\hat{\Phi}(\mathbf{r},t),\hat{\psi}(\mathbf{r}^{\prime},t)\right]=\left[\hat{\Phi}^{\dag}(\mathbf{r},t),\hat{\psi}^{\dag}(\mathbf{r}^{\prime},t)\right]=0\end{array}\label{eq:6}\end{equation}
and\begin{equation}
\begin{array}{l}
\left[\hat{\eta}(\mathbf{r},t),\hat{\psi}^{\dag}(\mathbf{r}^{\prime},t)\right]=\delta(\mathbf{r}-\mathbf{r}^{\prime})-\phi(\mathbf{r},t)\phi^{*}(\mathbf{r}^{\prime},t)\equiv Q(\mathbf{r},\mathbf{r}^{\prime},t),\\
\left[\hat{\eta}(\mathbf{r},t),\hat{\psi}(\mathbf{r}^{\prime},t)\right]=\left[\hat{\eta}^{\dag}(\mathbf{r},t),\hat{\psi}^{\dag}(\mathbf{r}^{\prime},t)\right]=0\end{array}\label{eq:7}\end{equation}
respectively. We use the Bogoliubov transformation\begin{equation}
\hat{\eta}(\mathbf{r},t)=\sum_{k}\left(u_{k}(\mathbf{r},t)\hat{a}_{k}+v_{k}^{*}(\mathbf{r},t)\hat{a}_{k}^{\dag}\right)\label{eq:8}\end{equation}
to diagonalize the grand-canonical Hamiltonian (as with standard HFB),
where we assume that the operators $\hat{a}_{k}$ obey the Bosonic
commutation relations\begin{equation}
\begin{array}{ccc}
\left[\hat{a}_{k},\hat{a}_{l}^{\dag}\right]=\delta_{kl} & \textrm{and} & \left[\hat{a}_{k},\hat{a}_{l}\right]=\left[\hat{a}_{k}^{\dag},\hat{a}_{l}^{\dag}\right]=0.\end{array}\label{eq:8a}\end{equation}
We use the respective Heisenberg equations of motion for the operators
$\hat{\Phi}$ and $\hat{\eta}$ and the commutation relations\begin{equation}
\begin{array}{lcl}
\left[\hat{\psi}(\mathbf{r},t),\hat{a}_{q}^{\dag}\right]=u_{q}(\mathbf{r},t), &  & \left[\hat{a}_{q},\hat{\psi}(\mathbf{r},t)\right]=v_{q}^{*}(\mathbf{r},t),\\
\left[\hat{\psi}^{\dag}(\mathbf{r},t),\hat{a}_{q}^{\dag}\right]=v_{q}(\mathbf{r},t), &  & \left[\hat{a}_{q},\hat{\psi}^{\dag}(\mathbf{r},t)\right]=u_{q}^{*}(\mathbf{r},t)\end{array}\label{eq:9}\end{equation}
for the quasi-particle annihilation and creation operators $\hat{a}_{q}$
and $\hat{a}_{q}^{\dag}$ to derive the equations\begin{equation}
\int d\mathbf{r}\phi^{*}\left[-i\hbar\frac{\partial\hat{\Phi}}{\partial t}+\left(\hat{h}_{\Omega}-\mu+g\hat{\psi}^{\dag}\hat{\psi}\right)\hat{\psi}\right]=0\label{eq:10}\end{equation}
for the condensate operator $\hat{\Phi}$, and\begin{equation}
i\hbar\frac{\partial u_{q}(\mathbf{r},t)}{\partial t}=\int d\mathbf{r}^{\prime}\left[\hat{L}(\mathbf{r},\mathbf{r}^{\prime},t)u_{q}(\mathbf{r}^{\prime},t)+\hat{M}(\mathbf{r},\mathbf{r}^{\prime},t)v_{q}(\mathbf{r}^{\prime},t)\right]\label{eq:11a}\end{equation}
and\begin{equation}
i\hbar\frac{\partial v_{q}^{*}(\mathbf{r},t)}{\partial t}=\int d\mathbf{r}^{\prime}\left[\hat{L}(\mathbf{r},\mathbf{r}^{\prime},t)v_{q}^{*}(\mathbf{r}^{\prime},t)+\hat{M}(\mathbf{r},\mathbf{r}^{\prime},t)u_{q}^{*}(\mathbf{r}^{\prime},t)\right]\label{eq:11b}\end{equation}
for the quasi-particle amplitudes $u_{q}$ and $v_{q}$, where we
have defined the operators\begin{equation}
\begin{array}{ccl}
\hat{L}(\mathbf{r},\mathbf{r}^{\prime},t) & \equiv & Q(\mathbf{r},\mathbf{r}^{\prime},t)\left(\hat{h}(\mathbf{r}^{\prime})-\mu+2g\hat{\psi}^{\dag}(\mathbf{r}^{\prime},t)\hat{\psi}(\mathbf{r}^{\prime},t)\right)\\
\hat{M}(\mathbf{r},\mathbf{r}^{\prime},t) & \equiv & Q(\mathbf{r},\mathbf{r}^{\prime},t)\left(g\hat{\psi}(\mathbf{r}^{\prime},t)\hat{\psi}(\mathbf{r}^{\prime},t)\right)\end{array}\label{eq:12}\end{equation}
for notational convenience. We note that the quantity in square brackets
in Eq. (\ref{eq:10}) is orthogonal to $\phi^{*}$. Hence we can re-write
(\ref{eq:10}) in the form\begin{equation}
i\hbar\frac{\partial\hat{\Phi}(\mathbf{r},t)}{\partial t}=\left(\hat{h}_{\Omega}-\mu+g\hat{\psi}^{\dag}(\mathbf{r},t)\hat{\psi}(\mathbf{r},t)\right)\hat{\psi}(\mathbf{r},t)-\hat{A}(\mathbf{r},t)\label{eq:13}\end{equation}
where $\hat{A}(\mathbf{r},t)$ is a quantity orthogonal to $\phi$,
i.e. $\int d\mathbf{r}\phi^{*}\hat{A}(\mathbf{r},t)=0$ chosen such
that angular and linear momentum conservation hold. Using the mean-field
approximations $\hat{a_{c}}\phi\rightarrow\Phi$, $\hat{\eta}^{\dag}\hat{\eta}\rightarrow\left\langle \hat{\eta}^{\dag}\hat{\eta}\right\rangle \equiv\tilde{n}$,
$\hat{\eta}\hat{\eta}\rightarrow\left\langle \hat{\eta}\hat{\eta}\right\rangle \equiv\tilde{m}$
, $\hat{\eta}^{\dag}\hat{\eta}^{\dag}\rightarrow\left\langle \hat{\eta}^{\dag}\hat{\eta}^{\dag}\right\rangle \equiv\tilde{m}^{*}$,
and $\hat{\psi}^{\dag}\hat{\psi}\hat{\psi}\rightarrow\left\langle \hat{\psi}^{\dag}\hat{\psi}\hat{\psi}\right\rangle =\left(\left|\Phi\right|^{2}+2\tilde{n}\right)\Phi+\tilde{m}\Phi^{*}$
and result (\ref{eq:14}) for $\hat{A}$ which we prove shortly, we
then obtain the orthogonal HFB equations consisting of the modified
generalized Gross-Pitaevskii equation (GGPE)\begin{equation}
\begin{array}{ccl}
i\hbar\frac{\partial}{\partial t}\Phi(\mathbf{r},t) & = & \left(\hat{h}(\mathbf{r})-\mu+g\left(\left|\Phi\right|^{2}+2\tilde{n}\right)\right)\Phi(\mathbf{r},t)\\
 &  & +g\tilde{m}\Phi^{*}(\mathbf{r},t)-\int d\mathbf{r}^{\prime}\hat{P}(\mathbf{r}^{\prime},\mathbf{r},t)\phi(\mathbf{r}^{\prime},t)\end{array}\label{eq:16}\end{equation}
and the orthogonal Bogoliubov-de Gennes equations (BdGEs)\begin{equation}
i\hbar\frac{\partial}{\partial t}\mathbf{w}_{q}(\mathbf{r},t)=\int d\mathbf{r}^{\prime}\hat{\mathbf{L}}(\mathbf{r},\mathbf{r}^{\prime},t)\mathbf{w}_{q}(\mathbf{r}^{\prime},t)\label{eq:17}\end{equation}
where we have defined the matrix operator\begin{equation}
\hat{\mathbf{L}}(\mathbf{r},\mathbf{r}^{\prime},t)\equiv\left[\begin{array}{cc}
\mathcal{\hat{L}}(\mathbf{r},\mathbf{r}^{\prime},t) & \hat{\mathcal{M}}(\mathbf{r},\mathbf{r}^{\prime},t)\\
-\hat{\mathcal{M}}^{*}(\mathbf{r},\mathbf{r}^{\prime},t) & -\mathcal{\hat{L}}^{*}(\mathbf{r},\mathbf{r}^{\prime},t)\end{array}\right]\label{eq:18}\end{equation}
and the vector\begin{equation}
\mathbf{w}_{q}(\mathbf{r},t)\equiv\left[\begin{array}{c}
u_{q}(\mathbf{r},t)\\
v_{q}(\mathbf{r},t)\end{array}\right]\label{eq:19}\end{equation}
for the quasi-particle amplitudes. The operators $\hat{P}(\mathbf{r}^{\prime},\mathbf{r},t)$,
$\mathcal{\hat{L}}(\mathbf{r},\mathbf{r}^{\prime},t)$ and $\hat{\mathcal{M}}(\mathbf{r},\mathbf{r}^{\prime},t)$
are defined by \begin{equation}
\hat{P}(\mathbf{r}^{\prime},\mathbf{r},t)\equiv\frac{\left(\tilde{n}(\mathbf{r}^{\prime},\mathbf{r},t)\mathcal{\hat{L}}(\mathbf{r}^{\prime},t)+\tilde{m}(\mathbf{r}^{\prime},\mathbf{r},t)\hat{\mathcal{M}}^{*}(\mathbf{r}^{\prime},t)\right)}{\sqrt{N_{c}(t)}}\label{eq:20}\end{equation}
and\begin{equation}
\begin{array}{ccl}
\mathcal{\hat{L}}(\mathbf{r},\mathbf{r}^{\prime},t) & \equiv & Q(\mathbf{r},\mathbf{r}^{\prime},t)\mathcal{\hat{L}}(\mathbf{r}^{\prime},t),\\
\mathcal{\hat{\mathcal{M}}}(\mathbf{r},\mathbf{r}^{\prime},t) & \equiv & Q(\mathbf{r},\mathbf{r}^{\prime},t)\mathcal{\hat{\mathcal{M}}}(\mathbf{r}^{\prime},t)\end{array}\label{eq:21}\end{equation}
where\begin{equation}
\begin{array}{ccl}
\mathcal{\hat{L}}(\mathbf{r},t) & \equiv & \hat{h}_{\Omega}(\mathbf{r})-\mu+2g\left(\left|\Phi(\mathbf{r},t)\right|^{2}+\tilde{n}(\mathbf{r},t)\right)\\
\hat{\mathcal{M}}(\mathbf{r},t) & \equiv & g\left(\Phi^{2}(\mathbf{r},t)+\tilde{m}(\mathbf{r},t)\right),\end{array}\label{eq:15}\end{equation}
with the definitions\begin{equation}
\begin{array}{ccl}
\tilde{n}(\mathbf{r}^{\prime},\mathbf{r},t) & \equiv & \left\langle \hat{\eta}^{\dag}(\mathbf{r}^{\prime},t)\hat{\eta}(\mathbf{r},t)\right\rangle \\
 & = & \sum_{q}\left[u_{q}^{*}(\mathbf{r}^{\prime},t)u_{q}(\mathbf{r},t)N_{BE}(\epsilon_{q})\right.\\
 &  & \left.+v_{q}(\mathbf{r}^{\prime},t)v_{q}^{*}(\mathbf{r},t)\left(N_{BE}(\epsilon_{q})+1\right)\right]\end{array}\label{eq:22}\end{equation}
 and\begin{equation}
\begin{array}{ccl}
\tilde{m}(\mathbf{r}^{\prime},\mathbf{r},t) & \equiv & \left\langle \hat{\eta}(\mathbf{r}^{\prime},t)\hat{\eta}(\mathbf{r},t)\right\rangle \\
 & = & \sum_{q}\left[v_{q}^{*}(\mathbf{r}^{\prime},t)u_{q}(\mathbf{r},t)N_{BE}(\epsilon_{q})\right.\\
 &  & \left.+u_{q}(\mathbf{r}^{\prime},t)v_{q}^{*}(\mathbf{r},t)\left(N_{BE}(\epsilon_{q})+1\right)\right]\end{array}\label{eq:23}\end{equation}
for the normal and anomalous correlation functions $\tilde{n}(\mathbf{r}^{\prime},\mathbf{r},t)$
and $\tilde{m}(\mathbf{r}^{\prime},\mathbf{r},t)$, where\begin{equation}
N_{BE}(\epsilon_{q})=\frac{1}{\exp(\beta\epsilon_{q})-1}\label{eq:23a1}\end{equation}
is the usual Bose distribution, with $\beta\equiv1/k_{B}T$ the temperature
parameter, and $k_{B}$ is the Boltzmann constant. Clearly $\tilde{n}(\mathbf{r},t)\equiv\tilde{n}(\mathbf{r},\mathbf{r},t)$
and $\tilde{m}(\mathbf{r},t)\equiv\tilde{m}(\mathbf{r},\mathbf{r},t)$
represent the usual thermal and anomalous densities. To prove Eq.
(\ref{eq:16}), we note using results from vector calculus, that we
can write \cite{key-10a} \[
\begin{array}{rl}
\frac{d}{dt}\left\langle \hat{\mathbf{L}}\right\rangle  & =\frac{d}{dt}\left\langle \int d\mathbf{r}\hat{\psi}^{\dag}(\mathbf{r},t)\hat{\mathbf{L}}\hat{\psi}(\mathbf{r},t)\right\rangle \\
 & =\frac{1}{i\hbar}\int d\mathbf{r}\left\{ \left|\Phi\right|^{2}\hat{\mathbf{L}}V_{T}+\left|\Phi\right|^{2}\hat{\mathbf{L}}\left(\left|\Phi\right|^{2}+2\tilde{n}\right)\right.\\
 & \left.+\tilde{n}\hat{\mathbf{L}}V_{T}+2\tilde{n}\hat{\mathbf{L}}\left(\left|\Phi\right|^{2}+\tilde{n}\right)\right.\\
 & \left.-\frac{1}{2}g\tilde{m}^{*}\hat{\mathbf{L}}\left(\Phi^{2}\right)-\frac{1}{2}g\tilde{m}\hat{\mathbf{L}}\left(\Phi^{*^{2}}\right)\right.\\
 & \left.-\frac{1}{2}g\left(\Phi^{*^{2}}+\tilde{m}^{*}\right)\hat{\mathbf{L}}\tilde{m}-\frac{1}{2}g\left(\Phi^{2}+\tilde{m}\right)\hat{\mathbf{L}}\tilde{m}^{*}\right.\\
 & \left.+\left\langle -\hat{A}^{\dag}+\frac{1}{\sqrt{N_{c}}}\hat{\eta}^{\dag}(\mathbf{r},t)\int d\mathbf{r}^{\prime}\hat{\eta}(\mathbf{r}^{\prime},t)\mathcal{\hat{L}}^{*}(\mathbf{r}^{\prime},t)\phi^{*}(\mathbf{r}^{\prime},t)\right.\right.\\
 & \left.\left.+\frac{1}{\sqrt{N_{c}}}\hat{\eta}^{\dag}(\mathbf{r},t)\int d\mathbf{r}^{\prime}\hat{\eta}^{\dag}(\mathbf{r}^{\prime},t)\hat{\mathcal{M}}(\mathbf{r}^{\prime},t)\phi^{*}(\mathbf{r}^{\prime},t)\right\rangle \hat{\mathbf{L}}\Phi\right.\\
 & \left.+\left\langle -\hat{A}+\frac{1}{\sqrt{N_{c}}}\left(\int d\mathbf{r}^{\prime}\hat{\eta}^{\dag}(\mathbf{r}^{\prime},t)\mathcal{\hat{L}}(\mathbf{r}^{\prime},t)\phi(\mathbf{r}^{\prime},t)\right)\hat{\eta}(\mathbf{r},t)\right.\right.\\
 & \left.\left.+\frac{1}{\sqrt{N_{c}}}\left(\int d\mathbf{r}^{\prime}\hat{\eta}(\mathbf{r}^{\prime},t)\hat{\mathcal{M}}^{*}(\mathbf{r}^{\prime},t)\phi(\mathbf{r}^{\prime},t)\right)\hat{\eta}(\mathbf{r},t)\right\rangle \hat{\mathbf{L}}\Phi^{*}\right\} \\
 & =\frac{1}{i\hbar}\int d\mathbf{r}\left(\left|\Phi\right|^{2}+\tilde{n}\right)\hat{\mathbf{L}}V_{T}.\end{array}\]
 provided we choose\begin{equation}
\hat{A}=\frac{\int d\mathbf{r}^{\prime}\left(\hat{\eta}^{\dag}(\mathbf{r}^{\prime},t)\mathcal{\hat{L}}(\mathbf{r}^{\prime},t)+\hat{\eta}(\mathbf{r}^{\prime},t)\hat{\mathcal{M}}{}^{*}(\mathbf{r}^{\prime},t)\right)\phi(\mathbf{r}^{\prime},t)\hat{\eta}(\mathbf{r},t)}{\sqrt{N_{c}(t)}},\label{eq:14}\end{equation}
thus ensuring that angular and linear momentum are conserved. Since
standard HFB is a $\phi$-derivable, and hence a conserving, theory,
particle number and both angular and linear momentum conservation
hold for standard HFB. It can be shown that particle number, and both
angular and linear momentum conservation also hold for this formalism,
the latter being ensured by the inclusion of the operator $\hat{P}$
in (\ref{eq:16}). The corresponding time-independent equations are
given by the modified GGPE\begin{equation}
\begin{array}{ccl}
\mu\Phi(\mathbf{r}) & = & \left(\hat{h}_{\Omega}(\mathbf{r})+g\left(\left|\Phi\right|^{2}+2\tilde{n}\right)\right)\Phi(\mathbf{r})\\
 &  & +g\tilde{m}\Phi^{*}(\mathbf{r})-\int d\mathbf{r}^{\prime}\hat{P}(\mathbf{r}^{\prime},\mathbf{r})\phi(\mathbf{r}^{\prime})\end{array}\label{eq:23a}\end{equation}
and the orthogonal BdGEs\begin{equation}
\epsilon_{q}\mathbf{w}_{q}(\mathbf{r})=\int d\mathbf{r}^{\prime}\hat{\mathbf{L}}(\mathbf{r},\mathbf{r}^{\prime},)\mathbf{w}_{q}(\mathbf{r}^{\prime})\label{eq:23b}\end{equation}
where the operators $\hat{\mathbf{L}}$ and $\hat{P}$ are now time
independent. To show the existence of a null subspace of zero-energy
eigenvalue modes for the orthogonal BdGEs (\ref{eq:23b}) spanned
by the mode $\left(\phi,-\phi^{*}\right)$, let us write $u_{0}(\mathbf{r})=\alpha\phi(\mathbf{r})$,
$v_{0}(\mathbf{r})=-\beta\phi^{*}(\mathbf{r})$ for some $\alpha,\beta\in\mathbb{C}$,
where $\phi(\mathbf{r})=\Phi(\mathbf{r})/\sqrt{N_{c}}$ , and $\phi(\mathbf{r})$
is normalized to unity, i.e.,\[
\int d\mathbf{r}\left|\phi(\mathbf{r})\right|^{2}=1.\]
Then, substituting $u_{0}$, $v_{0}$ into the first of the modified
time-independent BdGEs (\ref{eq:23b}), we find that\[
\begin{array}{ccl}
\epsilon_{0}\phi & = & \int d\mathbf{r}^{\prime}Q(\mathbf{r},\mathbf{r}^{\prime})\left[\alpha\mathcal{\hat{L}}(\mathbf{r}^{\prime},t)\phi(\mathbf{r}^{\prime})-\beta\mathcal{\hat{M}}(\mathbf{r}^{\prime},t)\phi^{*}(\mathbf{r}^{\prime})\right]\\
 & = & \alpha\mathcal{\hat{L}}(\mathbf{r}^{\prime},t)\phi(\mathbf{r}^{\prime})-\beta\mathcal{\hat{M}}(\mathbf{r}^{\prime},t)\phi^{*}(\mathbf{r}^{\prime})\\
 &  & -\phi(\mathbf{r})\int d\mathbf{r}^{\prime}\phi^{*}(\mathbf{r})\left[\alpha\mathcal{\hat{L}}(\mathbf{r}^{\prime},t)\phi(\mathbf{r}^{\prime})-\beta\mathcal{\hat{M}}(\mathbf{r}^{\prime},t)\phi^{*}(\mathbf{r}^{\prime})\right].\end{array}\]
Multiplying both sides by $\phi^{*}$, and integrating over all space,
we find {[}in view of the orthonormality of $\phi$ and of the orthogonality
condition (\ref{eq:5})] that\[
\begin{array}{ccl}
\epsilon_{0} & = & \phi(\mathbf{r})\int d\mathbf{r}^{\prime}\phi^{*}(\mathbf{r})\left[\alpha\mathcal{\hat{L}}(\mathbf{r}^{\prime},t)\phi(\mathbf{r}^{\prime})-\beta\mathcal{\hat{M}}(\mathbf{r}^{\prime},t)\phi^{*}(\mathbf{r}^{\prime})\right]\\
 &  & -\phi(\mathbf{r})\int d\mathbf{r}^{\prime}\phi^{*}(\mathbf{r})\left[\alpha\mathcal{\hat{L}}(\mathbf{r}^{\prime},t)\phi(\mathbf{r}^{\prime})-\beta\mathcal{\hat{M}}(\mathbf{r}^{\prime},t)\phi^{*}(\mathbf{r}^{\prime})\right]\\
 & = & 0.\end{array}\]
We have thus shown in the time-independent case for this formalism
that there exists a zero-energy eigenvalue. However, the existence
of a zero-energy eigenvalue does not imply that the Pine-Hugenholtz
theorem \cite{key-11} is satisfied or indeed that this corresponds
to the Goldstone mode, nor is the remainder of the energy spectrum
{}``corrected,'' and is quite similar to the standard HFB spectrum
in spite of the existence of a zero-energy eigenvalue.

We now obtain from the modified GGPE (\ref{eq:16}), the continuity
equation for the condensate density\begin{equation}
\frac{\partial}{\partial t}\left|\Phi(\mathbf{r},t)\right|^{2}+\mathbf{\nabla}.\mathbf{j}(\mathbf{r},t)=\mathbf{\Omega}.\left(\mathbf{r}\times\mathbf{\nabla}\right)\left|\Phi(\mathbf{r},t)\right|^{2}-\frac{i}{\hbar}C(\mathbf{r},t)\label{eq:24}\end{equation}
 where\begin{equation}
\mathbf{j}(\mathbf{r},t)\equiv\frac{i\hbar}{2m}\left(\Phi(\mathbf{r},t)\mathbf{\nabla}\Phi^{*}(\mathbf{r},t)-\Phi^{*}(\mathbf{r},t)\mathbf{\nabla}\Phi(\mathbf{r},t)\right)\label{eq:24a}\end{equation}
is the current density, and where we have defined the quantity\begin{equation}
\begin{array}{ccl}
C(\mathbf{r},t) & = & g\left(\tilde{m}(\mathbf{r},t)\Phi^{*^{2}}(\mathbf{r},t)-\tilde{m}^{*}(\mathbf{r},t)\Phi^{2}(\mathbf{r},t)\right)\\
 &  & +G^{*}(\mathbf{r},t)-G(\mathbf{r},t)\end{array}\label{eq:25}\end{equation}
with\begin{equation}
G(\mathbf{r},t)\equiv\Phi^{*}(\mathbf{r},t)\int d\mathbf{r}^{\prime}\hat{P}(\mathbf{r}^{\prime},\mathbf{r},t)\phi(\mathbf{r}^{\prime},t),\label{eq:26}\end{equation}
and use this to find an expression for the precessional frequencies
of off-axis vortices and vortex arrays in quasi-two-dimensional BECs.
To do this we consider a stationary vortex system in the frame rotating
at the precessional frequency of the vortex or vortices.

\section{Calculation of Precessional Frequencies of Vortices in Quasi-two-dimensional
BECs}

In the quasi-2D regime, the time-independent orthogonal HFB equations
in a frame rotating with angular frequency $\Omega$ are given in
polar coordinates by the time-independent GGPE\begin{equation}
\begin{array}{ccl}
\mu\Phi & = & \left(\hat{h}_{\Omega}+C_{2D}\left(\left|\Phi\right|^{2}+2\tilde{n}\right)\right)\Phi+C_{2D}\tilde{m}\Phi^{*}\\
 &  & -\int_{0}^{2\pi}\int_{0}^{\infty}r^{\prime}dr^{\prime}d\theta^{\prime}\hat{P}(r^{\prime},\theta^{\prime},r,\theta)\phi(r^{\prime},\theta^{\prime})\end{array}\label{eq:27}\end{equation}
and the 2D time-independent Bogoliubov de Gennes Equations (BdGEs)\begin{equation}
\begin{array}{ccl}
\epsilon_{q}\mathbf{w}_{q}(r,\theta) & = & \int_{0}^{2\pi}\int_{0}^{\infty}r^{\prime}dr^{\prime}d\theta^{\prime}\hat{\mathbf{L}}(r,\theta,r^{\prime},\theta^{\prime})\mathbf{w}_{q}(r^{\prime},\theta^{\prime})\end{array}\label{eq:28}\end{equation}
where\begin{equation}
\mathbf{w}_{q}(r,\theta)\equiv\left[\begin{array}{c}
u_{q}(r,\theta)\\
v_{q}(r,\theta)\end{array}\right]\label{eq:29}\end{equation}
and\begin{equation}
\hat{\mathbf{L}}(r,\theta,r^{\prime},\theta^{\prime})\equiv\left[\begin{array}{cc}
\mathcal{\hat{L}}(r,\theta,r^{\prime},\theta^{\prime}) & \hat{\mathcal{M}}(r,\theta,r^{\prime},\theta^{\prime})\\
-\hat{\mathcal{M}}^{*}(r,\theta,r^{\prime},\theta^{\prime}) & -\mathcal{\hat{L}}^{*}(r,\theta,r^{\prime},\theta^{\prime})\end{array}\right].\label{eq:30}\end{equation}
Here we have defined the operators\begin{equation}
\begin{array}{ccl}
\mathcal{\hat{L}}(r,\theta,r^{\prime},\theta^{\prime}) & \equiv & Q(r,\theta,r^{\prime},\theta^{\prime})\mathcal{\hat{L}}(r^{\prime},\theta^{\prime})\\
\hat{\mathcal{M}}(r,\theta,r^{\prime},\theta^{\prime}) & \equiv & Q(r,\theta,r^{\prime},\theta^{\prime})\hat{\mathcal{M}}(r^{\prime},\theta^{\prime})\end{array}\label{eq:31}\end{equation}
and\begin{equation}
\begin{array}{ccl}
\hat{P}(r^{\prime},\theta^{\prime},r,\theta) & \equiv & \left(\tilde{n}(r^{\prime},\theta^{\prime},r,\theta)\mathcal{\hat{L}}(r^{\prime},\theta^{\prime})\right.\\
 &  & \left.+\tilde{m}(r^{\prime},\theta^{\prime},r,\theta)\hat{\mathcal{M}}^{*}(r^{\prime},\theta^{\prime})\right)/\sqrt{N_{c}(t)}\end{array}\label{eq:32}\end{equation}
with\begin{equation}
\begin{array}{ccl}
\mathcal{\hat{L}}(r,\theta) & \equiv & \hat{h}_{\Omega}(r,\theta)-\mu+2C_{2D}\left(\left|\Phi(r,\theta)\right|^{2}+\tilde{n}(r,\theta)\right)\\
\hat{\mathcal{M}}(r,\theta) & \equiv & C_{2D}\left(\Phi^{2}(r,\theta)+\tilde{m}(r,\theta)\right),\end{array}\label{eq:33}\end{equation}
\begin{equation}
Q(r,\theta,r^{\prime},\theta^{\prime})\equiv\frac{1}{r^{\prime}}\delta(r-r^{\prime})\delta(\theta-\theta^{\prime})-\phi(r,\theta)\phi^{*}(r^{\prime},\theta^{\prime}),\label{eq:33a}\end{equation}
and\begin{equation}
\hat{h}_{\Omega}(r,\theta)=-\left(\frac{\partial^{2}}{\partial r^{2}}+\frac{1}{r}\frac{\partial}{\partial r}+\frac{1}{r^{2}}\frac{\partial^{2}}{\partial\theta^{2}}\right)+i\Omega\frac{\partial}{\partial\theta}+r^{2}.\label{eq:34}\end{equation}
We note that in the quasi-2D regime that $\mathbf{\Omega}$ reduces
to the component $\Omega\equiv\Omega_{z}$ along the $z$-axis. We
use the dimensionless units $r_{0}=\sqrt{\frac{\hbar}{m\omega_{r}}}$
for the length scale, and $t_{0}=\frac{2}{\omega_{r}}$ for the time
scale, and express the condensate wave function and quasiparticle
amplitudes in units of $\sqrt{N/r_{0}^{3}}$. Then the energies are
given in harmonic oscillator units $\hbar\omega_{r}/2$. The nonlinearity
constant $C_{2D}$ is found by expressing the interaction term $g$
in dimensionless units, and integrating over all $z$ yielding $C_{2D}=8\pi\left(\frac{\lambda}{2\pi}\right)^{1/2}\frac{a_{s}}{r_{0}}N$
where $N$ is the number of atoms, $a_{s}$ is the scattering length,
and $\lambda\equiv\omega_{z}/\omega_{r}$ is the trap aspect ratio.

Let $R(\mathbf{r})\equiv\textrm{Re}(\Phi(\mathbf{r}))$ and $I(\mathbf{r})\equiv\textrm{Im}(\Phi(\mathbf{r}))$.
Then in the two-dimensional system, the continuity equation for the
condensate density is given by\begin{equation}
\Omega r^{2}A(r,\theta)=B(r,\theta)\label{eq:35}\end{equation}
where we have defined\begin{equation}
A(r,\theta)\equiv\left(R\frac{\partial R}{\partial\theta}+I\frac{\partial I}{\partial\theta}\right),\label{eq:36}\end{equation}
and\begin{equation}
\begin{array}{ccl}
B(r,\theta) & \equiv & \left[r^{2}\left(R\frac{\partial^{2}I}{\partial r^{2}}-I\frac{\partial^{2}R}{\partial r^{2}}\right)+r\left(R\frac{\partial I}{\partial r}-I\frac{\partial R}{\partial r}\right)\right.\\
 &  & \left.+\left(R\frac{\partial^{2}I}{\partial\theta^{2}}-I\frac{\partial^{2}R}{\partial\theta^{2}}\right)\right]+i\frac{r^{2}}{2}C(r,\theta).\end{array}\label{eq:37}\end{equation}
with\begin{equation}
\begin{array}{ccl}
C(r,\theta) & \equiv & C_{2D}\left(\tilde{m}(r,\theta)\Phi^{*^{2}}(r,\theta)-\tilde{m}^{*}(r,\theta)\Phi^{2}(r,\theta)\right)\\
 &  & +G^{*}(r,\theta)-G(r,\theta)\end{array}\label{eq:38}\end{equation}
and\begin{equation}
G(r,\theta)\equiv\Phi^{*}(r,\theta)\int_{0}^{2\pi}\int_{0}^{\infty}r^{\prime}dr^{\prime}d\theta^{\prime}\hat{P}(r^{\prime},\theta^{\prime},r,\theta)\phi(r^{\prime},\theta^{\prime}).\label{eq:39}\end{equation}
Multiplying both sides by $A(r,\theta)$, and integrating over all
space gives us the expression\begin{equation}
\Omega=\frac{\int_{0}^{2\pi}\int_{0}^{\infty}A(r,\theta)B(r,\theta)rdrd\theta}{\int_{0}^{2\pi}\int_{0}^{\infty}r^{2}\left[A(r,\theta)\right]^{2}rdrd\theta}\label{eq:40}\end{equation}
which we solve self-consistently with the time-independent orthogonal
HFB equations in the frame rotating at angular frequency $\Omega$.
The converged solution then represents a stationary solution in the
rotating frame, and $\Omega$ corresponds to the precessional frequency
of the off-axis vortex or vortex array.

A vortex at position $\left(r_{1},\theta_{1}\right)$ may be created
by expanding the condensate wave function in terms of modified Laguerre
basis functions $\left\{ \chi_{ln}^{(1)}(r,\theta)\right\} $ as follows:

Define\begin{equation}
\mathcal{S}^{(1)}\equiv\left\{ l,n\mid\left(l,n\right)\in\mathcal{S}-\left\{ \left(1,0\right)\right\} \right\} \label{eq:40a}\end{equation}
where $\mathcal{S}\equiv\left\{ l,n\mid n=0,\ldots,l=0,\pm1,\ldots\right\} $,
then\begin{equation}
\Phi(r,\theta)=\sum_{l,n\in\mathcal{S}^{(1)}}a_{ln}\chi_{ln}^{(1)}(r,\theta)\label{eq:41}\end{equation}
provided $\left(r_{1},\theta_{1}\right)$ is not a root of $\xi_{ln}(r,\theta)$,
where we define\begin{equation}
\chi_{ln}^{(1)}(r,\theta)\equiv\xi_{ln}(r,\theta)-\frac{\xi_{ln}(r_{1},\theta_{1})}{\xi_{10}(r_{1},\theta_{1})}\xi_{10}(r,\theta),\label{eq:42}\end{equation}
with\begin{equation}
\xi_{ln}(r,\theta)=\frac{e^{il\theta}}{\sqrt{2\pi}}\left(\frac{2n!}{(n+\left|l\right|)!}\right)^{1/2}e^{-r^{2}/2}r^{\left|l\right|}L_{n}^{\left|l\right|}(r^{2}),\label{eq:43}\end{equation}
and $L_{n}^{\left|l\right|}(x)$ a modified Laguerre polynomial of
order $n$. The superscript in the representation of the modified
basis function, indicates a single vortex. This is motivated simply
by expanding the condensate wave function in terms of the complete
Laguerre basis $\left\{ \xi_{ln}(r,\theta)\mid l,n\in\mathcal{S}\right\} $
and considering that $\Phi(r_{1},\theta_{1})=\sum_{ln}a_{ln}\xi_{ln}(r_{1},\theta_{1})=0$.
Imposing this condition implies dependency of one of the coefficients
$a_{ln}$ on the others, reducing the dimension of the vector space
by one. Here we choose to represent $a_{10}$ in terms of the remainder
$\left\{ a_{ln}\mid l,n\in\mathcal{S}^{(1)}\right\} $. We find that
$a_{10}=-\sum_{l,n\in\mathcal{S}^{(1)}}\left(\xi_{ln}(r_{1},\theta_{1})/\xi_{10}(r_{1},\theta_{1})\right)a_{ln}$,
hence we can write $\Phi(r,\theta)$ in the form (\ref{eq:41}).

This procedure may be extended to $N_{v}$ vortices $\left\{ \left(r_{1},\theta_{1}\right),\ldots,\left(r_{N_{v}},\theta_{N_{v}}\right)\right\} $
by an iterative process, writing\begin{equation}
\Phi(r,\theta)=\sum_{l,n\in\mathcal{S}^{(N_{v})}}a_{ln}\chi_{ln}^{(N_{v})}(r,\theta)\label{eq:44}\end{equation}
where we have defined\begin{equation}
\chi_{ln}^{(N_{v})}(r,\theta)\equiv\chi_{ln}^{(N_{v}-1)}(r,\theta)-\frac{\chi_{ln}^{(N_{v}-1)}(r_{N_{v}},\theta_{N_{v}})}{\chi_{N_{v}0}^{(N_{v}-1)}(r_{N_{v}},\theta_{N_{v}})}\chi_{N_{v}0}^{(N_{v})}(r,\theta).\label{eq:45}\end{equation}
and where we have generalized the definition (\ref{eq:40a}) in the
single-vortex case to\begin{equation}
\mathcal{S}^{(N_{v})}\equiv\left\{ l,n\mid\left(l,n\right)\in\mathcal{S}-\left\{ \left(1,0\right),\ldots,\left(N_{v},0\right)\right\} \right\} \label{eq:45a}\end{equation}
in the multi-vortex case.

We illustrate the method by considering a dilute Bose gas consisting
of 2000 $^{87}\textrm{Rb }$atoms in an axially symmetric harmonic
trap, with radial and axial harmonic trapping frequencies of $\Omega_{r}=2\pi\times10\textrm{Hz}$
and $\Omega_{z}=2\pi\times400\textrm{Hz}$, respectively. Thus the
axial confinement is sufficiently strong that all excited axial states
may be neglected, hence the axial ($z$) component can then be integrated
out, and the BEC may effectively be treated as two-dimensional. In
order for this quasi-two-dimensional approximation to hold, we require
that the axial harmonic transition energy $\hbar\omega_{z}$ be much
greater than the temperature $T$ and the mean-field energy of the
BEC \cite{key-11a}. Here the axial harmonic energy transition corresponds
to a temperature of $T_{\Delta_{z}}=\hbar\omega_{z}/k_{B}=19.2\textrm{nK}$,
and the mean-field energy to an effective temperature of $T{}_{MF}=C_{2D}n_{0}^{(2D)}/k_{B}^{\prime}=.84\textrm{nK}$,
where $k_{B}$ is Boltzmann's constant, and $k_{B}^{\prime}=2k_{B}/\hbar\omega_{r}.$
In all these simulations the BEC temperature $T\leq10\textrm{nK}$,
so this approximation is justified.

Stationary solutions are found in the case of a single off-axis vortex,
and in the cases of triangular and hexagonal vortex arrays, provided
the interactions between the votices are not too strong. In cases
where there are strong interactions between vortices, approximately
stationary solutions can be found, establishing lower and upper bounds
for the precessional frequency, and the precessional frequency determined
more accurately in time-dependent simulations.

Precessional frequencies as a function of vortex position $a$ and
of temperature $T$ for the single off-axis case are presented in
Figs. 1(c) and (d) of a previous publication \cite{key-12}, and will
not be repeated here. Instead, we concentrate on the cases of triangular
and hexagonal vortex arrays.%
\begin{figure}
\includegraphics[scale=0.3]{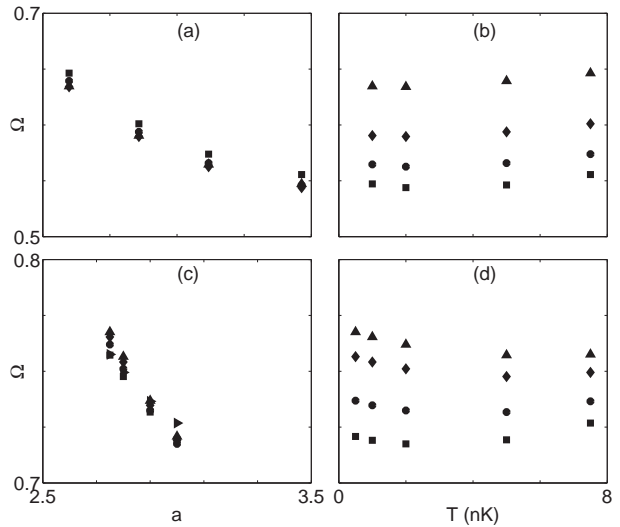}\caption{Triangular vortex array. (a) Precessional frequency $\Omega$ versus
lattice parameter $a$ at temperatures 1, 2, 5 and 7.5 nK (triangles,
diamonds, circles, squares); (b) precessional frequency $\Omega$
versus temperature $T$ for lattice parameter 1.5$\sqrt{3}$, 1.65$\sqrt{3}$,
1.8$\sqrt{3}$, and 2$\sqrt{3}$ (triangles, diamonds, circles, squares).
Hexagonal vortex array. (c) Precessional frequency $\Omega$ (in units
of radial trapping frequency $\omega_{r}$) versus lattice parameter
$a$ at temperatures 1, 2, 5 and 7.5 nK (triangles, diamonds, circles,
squares); (d) precessional frequency $\Omega$ versus temperature
$T$ for lattice parameter 2.75, 2.8, 2.9 and 3 (triangles, diamonds,
circles, squares). All frequencies are in units of the radial trapping
frequency $\omega_{r}$. All positions are in units of the harmonic
oscillator length $r_{0}$.}

\end{figure}
 The precessional frequency $\Omega$ as a function of temperature
$T$ for various values of lattice parameter $a$ and of lattice parameter
$a$ at various temperatures are shown in Figs. 1(a) and 1(b) for
the triangular vortex array and in Figs. 1(c) and 1(d) for the hexagonal
vortex array. We see that $\Omega$ decreases with increasing lattice
parameter $a$ over the range shown (within the Thomas-Fermi radius).
This is in accord with the areal density law given by the approximation
\cite{key-6a}\begin{equation}
\Omega=\frac{\hbar\pi}{m}\left[n_{v}+\frac{1}{2\pi}\frac{R_{TF}^{2}}{\left(R_{TF}^{2}-r^{2}\right)^{2}}\ln\left(e^{-1}\frac{\omega_{r}}{\Omega}\right)\right]\label{eq:45b}\end{equation}
where $n_{v}$ is the areal density of vortices approximated by $n_{v}=N_{v}/(\pi R_{TF}^{2})$,
$N_{v}$ is the number of vortices in the vortex array, and $R_{TF}$
is the Thomas-Fermi radius and may be understood by considering that
the vortex precession velocity is determined primarily by the background
velocity field around the vortex core, and the core shape \cite{key-12a,key-12b}.
The effect of their image charges in the condensate boundary is less
important, in contrast with the single vortex case \cite{key-12c,key-12d}.
In the single off-axis case $\Omega$ increases with increasing $a$
\cite{key-12}. This is in agreement with GPE results \cite{key-12c,key-12d}
and may be explained in terms of the effect of the vortex's image
charge in the boundary of the condensate \cite{key-12c,key-12d}.

Figures 2(a) and 2(b) show, respectively, simulated absorption images
for the condensate density and the thermal density \cite{key-12e},
and Fig. 2(c) the condensate phase in the $xy$-plane for a single
off-axis vortex situated at $a=0.5$. We note the phase singularity
in Fig. 2(c) indicating a singly charged vortex. Similarly Figs. 2(d)
and 2(e) show, respectively, simulated absorption images for the condensate
density and the thermal density, and Fig. 2(f) shows the condensate
phase in the $xy$-plane for a triangular vortex latice having lattice
parameter $a=1.65\sqrt{3}$. Figs. 2(g)-2(i) show the respective plots
for a hexagonal vortex latice having lattice parameter $a=2.9$. The
phase singularities in Figs. 2(f) and 2(i) indicate respectively,
three and seven singly charged vortices. We note from the simulated
absorption images for the condensate in Figs. 2(a), 2(d) and 2(g)
that the condensate extends further outward for the triangular and
hexagonal vortex arrays (i.e., larger Thomas-Fermi radius) than for
the single vortex case. This is consistent with our findings concerning
the dependency of $\Omega$ on $a$, where we find that $\Omega$
decreases with increasing $a$ in the vortex array case, in contrast
to the single off-axis vortex case. Unfortunately, it is not possible
to find stationary solutions in the limit of sufficiently large numbers
of vortices to validate against the areal density approximation \cite{key-6,key-7};
nevertheless we perform  this calculation for the case of seven vortices.%
\begin{figure}
\includegraphics[scale=0.29]{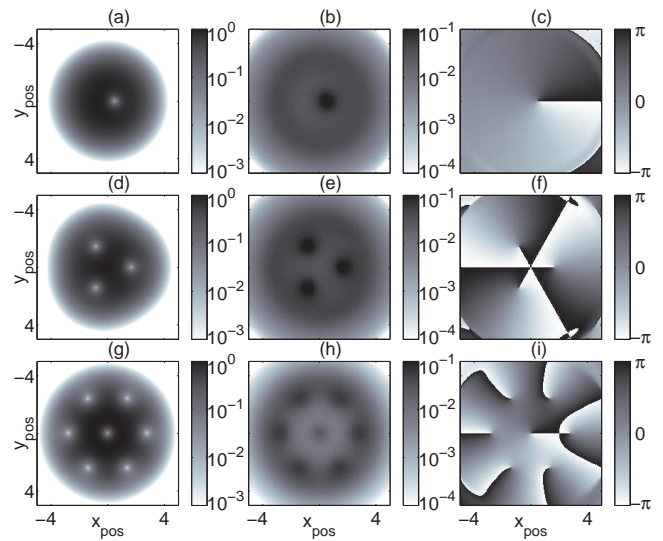}\caption{Simulated absorption images \cite{key-12e} for off-axis vortex, triangular
vortex array and hexagonal vortex array showing (a) condensate density,
(b) thermal density, (c) condensate phase for a single vortex at $a=\textrm{0.5}$,
and (d) condensate density, (e) thermal density, (f) condensate phase
for a triangular vortex array with lattice parameter $a=\textrm{1.65}\sqrt{3}$,
and (g) condensate density, (h) thermal density, (i) condensate phase
for a hexagonal vortex array with lattice parameter $a=$2.9. All
positions are in units of the harmonic oscillator length $r_{0}$.}

\end{figure}

Figures 3(a) and 3(b) represent plots of precessional frequency $\Omega$
versus lattice parameter $a$ at temperatures $T=2$ nK and $T=5$
nK, respectively, for vortex arrays composed of two, three,  and seven
vortices. We see that $\Omega$ increases with an increasing number
of vortices. The inserts of Figs. 3(a) and 3(b) seem to indicate that
$\Omega$ tends asymptotically to an upper limit as the number of
vortices increases, presumably the limit predicted by the areal density
approximation (\ref{eq:45b}). In the dimensionless units used here,
this may be written\begin{equation}
\Omega=\frac{1}{R_{TF}^{2}}\left[N_{v}+\frac{1}{2\left(1-\frac{r^{2}}{R_{TF}^{2}}\right)^{2}}\ln\left(e^{-1}\frac{\omega_{r}}{\Omega}\right)\right].\label{eq:45c}\end{equation}
 Therefore, for the hexagonal array where $N_{v}=7$, and taking $R_{TF}\sim3.5r_{0}$,
we predict to first approximation a precessional frequency of $\Omega\sim.57\omega_{r}$.
Taking into account the second term in (\ref{eq:45c}) with $r\sim0.5R_{TF}$,
we predict $\Omega\sim.55\omega_{r}$ (so the second term is unimportant
in the regime considered here). This is of the order of $20\%$ lower
than the value predicted using the continuity equation calculation.
We note that the areal density approximation is valid only for a large
number of vortices. Furthermore, referring to the inserts of Figs.
3(a) and (b) reveals that the precessional frequency approaches an
asymptotic value of $\sim.72$. We conclude, however, that these results
are consistent in the asymptotic limit.%
\begin{figure}
\includegraphics[scale=0.3]{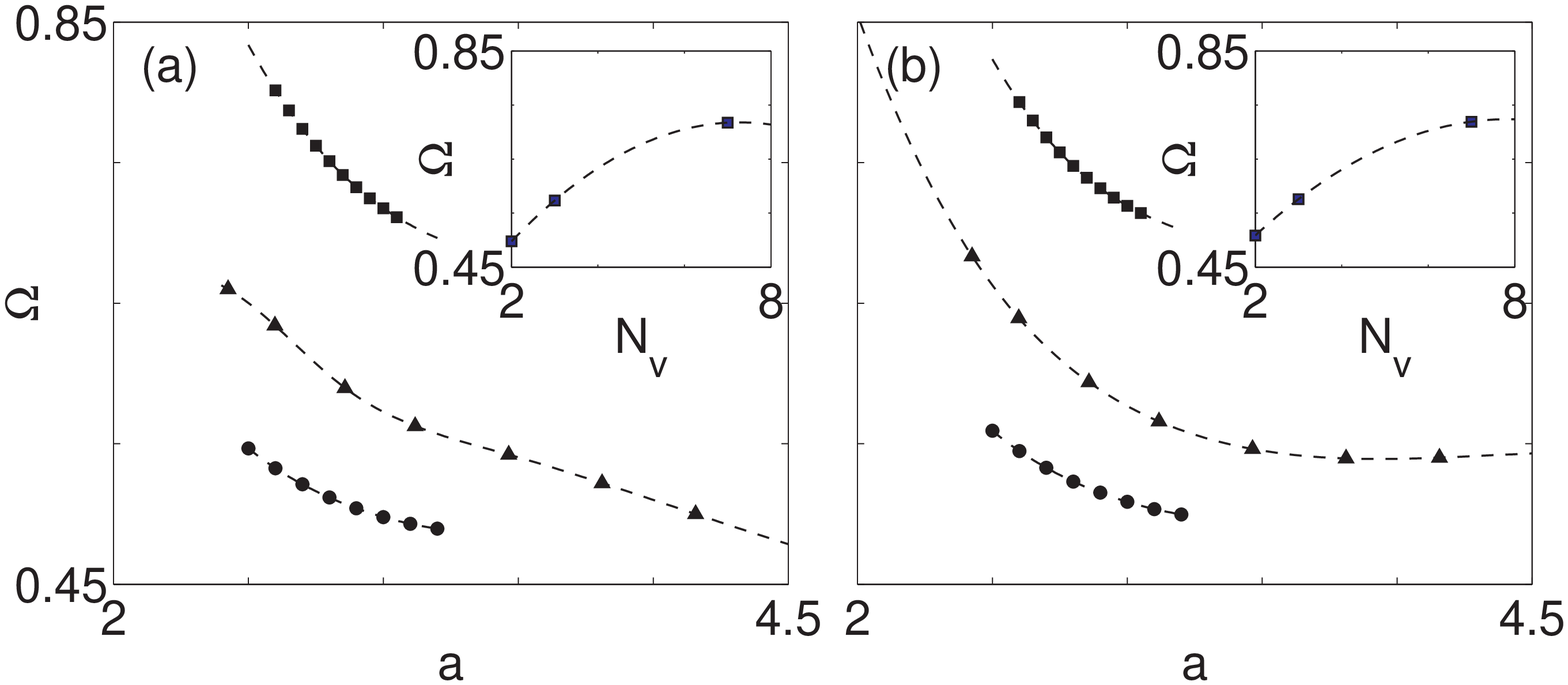}\caption{Precessional frequency $\Omega$ versus lattice parameter $a$ for
vortex arrays composed of two (circles), three (triangles), and seven
vortices (squares) with cubic spline interpolation (dashed line) (a)
at $T=2$ nK, and (b) at $T=5$ nK. The insert in each respective
figure shows the precessional frequency $\Omega$ versus number of
vortices for vortex arrays composed of two, three, and seven vortices
having lattice parameter $a=3.0$ with cubic spline interpolation
(dashed line). All frequencies are in units of the radial trapping
frequency $\omega_{r}$. All positions are in units of the harmonic
oscillator length $r_{0}$.}

\end{figure}

\section{Evolution of Time-independent Solutions}

Time-dependent simulations using the time-independent data initial
state are in good agreement with the above calculations. For a single
vortex at $a=0.5$ radial harmonic oscillator units from the axis,
the precessional frequency was estimated using a least-squares fit
of a sinusoid to the vortex $x$-displacement versus time and was
found to be $\Omega_{LS}=0.3794\omega_{r}$, in good agreement with
the value of $\Omega=0.3727\omega_{r}$, as predicted in the time-independent
calculations. The error in the prediction scales as the number of
computational basis functions which, for practical reasons, is only
209 in these calculations. For 839 computational basis functions,
the time-independent calculations predict a value of $\Omega=0.3761\omega_{r}$.
We see no evidence of dissipation during the time of the simulation
of five trap cycles, and the results reveal that the trajectory of
the vortex constitutes a circle.  

Time-dependent simulations were also performed at $T=5\textrm{ nK}$
for a symmetrical triangular vortex array with lattice parameter $a=1.65\sqrt{3}$
(i.e., three vortices symmetrically positioned at $1.65$ radial harmonic
oscillator units from the axis), and for a symmetrical hexagonal triangular
vortex centered on the axis having lattice parameter $a=2.85$ (i.e.,
seven vortices), again using time-independent calculations for the
initial state. In both simulations, the off-axis vortices precess
in a circle, and again we see no evidence of dissipation during the
time of the simulation of five trap cycles. The precessional frequency
$\Omega_{LS}=0.5856\omega_{r}$ for the triangular vortex array was
estimated using a least squares fit as before, in good agreement with
the value of $\Omega=0.5938\omega_{r}$, as predicted in the time-independent
calculations for 209 computational basis functions.

\section{Stirring of the Condensate}

In these, and in other time-dependent simulations, the temperature
$T$ remains fixed. It is not necessary to vary $\mu$ to maintain
particle conservation since the rate of change in the condensate population
$dN_{c}/dt$ and in the non-condensate population $d\tilde{N}/dt$
are equal and opposite, depending only on $\tilde{m}$ and $\Phi$,
i.e., $dN_{c}/dt=-d\tilde{N}/dt=i\frac{g}{\hbar}\int d\mathbf{r}\left(\tilde{m}\Phi^{*^{2}}-\tilde{m}^{*}\Phi^{2}\right)$,
therefore particle conservation always holds. In these simulations
we use one or more Gaussian rotating potentials in order to stir the
condensate. The stirring imparts angular momentum to the condensate,
and therefore one would expect the production of vortices in cases
where the angular frequency of the stirrer(s) exceeds a certain critical
value. This should correspond (at least approximately) to the velocity
of the stirrer with respect to the fluid exceeding the local Landau
critical velocity (i.e. the local speed of sound). In what follows
we shall establish a critical stirring frequency, below which no vortices
are produced in regions of appreciable density (i.e., within the Thomas-Fermi
radius). In these simulations we find that this critical frequency
corresponds to a velocity of the stirrer through the BEC that is slightly
in excess of the local Landau critical velocity (see Table 1). In
the simulations where the critical stirring frequency is exceeded,
we are able to determine by the method of least squares, as alluded
to above, the precessional frequencies and compare these with the
values predicted in the time-independent calculations using the continuity
equation for the condensate density. We shall see presently that the
agreement between the time-dependent values and the predicted values
using the time-independent calculations is good, as one would expect
in the adiabatic case since then the term $i\hbar\frac{\partial}{\partial t}\left|\Phi\right|^{2}$
in the time-dependent continuity equation for the condensate density
 (\ref{eq:24}) would be small, i.e., of order of the fluctuations
in the chemical potential. We stir the BEC using a single Gaussian
stirrer at $a$= $\textrm{1.5}r_{0}$ off axis, of amplitude 10 $\hbar\omega_{r}/2$,
full width at half maximum (FWHM) 0.82 $r_{0}$. Two cases are considered.
In the first we stir the BEC at a sub-critical stirring frequency
$\textrm{0.38 }\omega_{r}$, with the stirrer on for 10 trap cycles.
In the second case we stir the BEC with stirring frequency $\Omega=0.5\omega_{r}$
, with the stirrer on for 4.5 trap cycles. In both cases the stirring
frequency is ramped up from zero to the specified value over 0.2 trap
cycles. In the second case two vortices are produced, and the simulated
absorption images are shown in Fig. 4, and the trajectories of these
vortices in Fig. 5 \cite{key-12d1}.%
\begin{figure}
\includegraphics[scale=0.35]{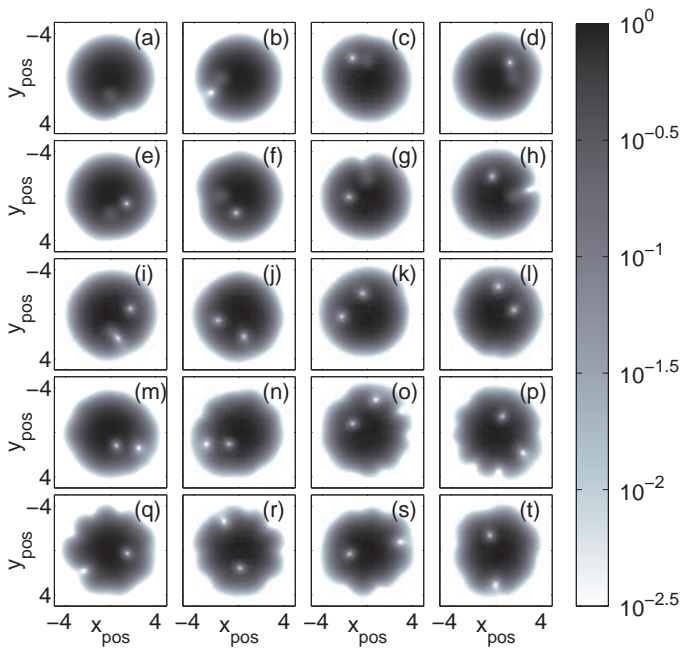}\caption{Simulated absorption images \cite{key-12e}; stirring of condensate
in an counterclockwise direction using one Gaussian stirrer at $a$=
$\textrm{1.5 }r_{0}$ off axis, of amplitude 10 $\hbar\omega_{r}/2$,
FWHM 0.82 $r_{0}$, switched on adiabatically over 0.2 trap cycles,
stirring frequency 0.5 $\omega_{r}$, stirrer on for 4.5 trap cycles
after (a) 0.5, (b) 1, (c) 1.5, (d) 2, (e) 2.5, (f) 3, (g) 3.5, (h)
4, (i) 4.5, (j) 5, (k) 5.5, (l) 6, (m) 6.5, (n) 7, (o) 7.5, (p) 8,
(q) 8.5, (r) 9, (s) 9.5, and (t) 10 trap cycles. All positions are
in units of the harmonic oscillator length $r_{0}$.}

\end{figure}
\begin{figure}
\includegraphics[scale=0.32]{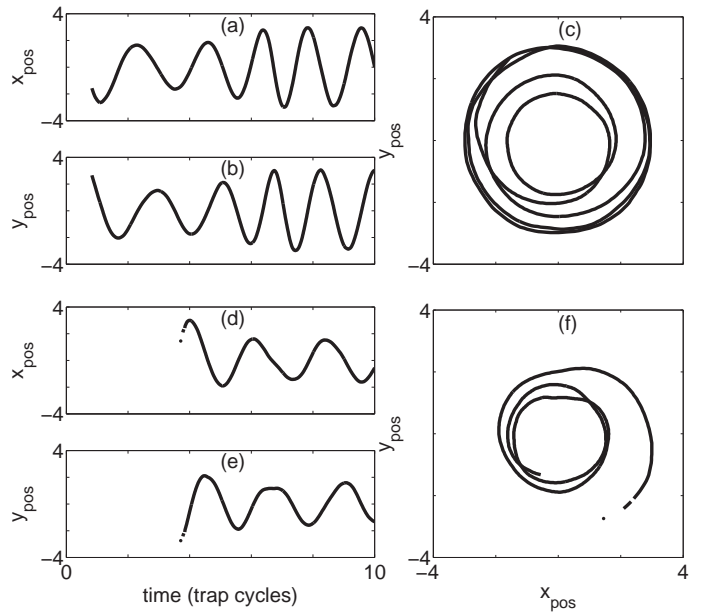}\caption{Stirring of condensate using one Gaussian stirrer at $a$= $\textrm{1.5}r_{0}$
off axis, of amplitude 10$\hbar\omega_{r}/2$, FWHM 0.82$r_{0}$,
switched on adiabatically over 0.2 trap cycles, stirring frequency
0.5$\omega_{r}$, stirrer on for 4.5 trap cycles (a) $x$-displacement
of vortex 1, (b) $y$-displacement of vortex 1, and (c) trajectory
of vortex 1 over a period of 10 trap cycles (d) $x$-displacement
of vortex 2, (e) $y$-displacement of vortex 2, and (f) trajectory
of vortex 2 over a period of 10 trap cycles. All positions are in
units of the harmonic oscillator length $r_{0}$.}

\end{figure}
\begin{figure}
\includegraphics[scale=0.3]{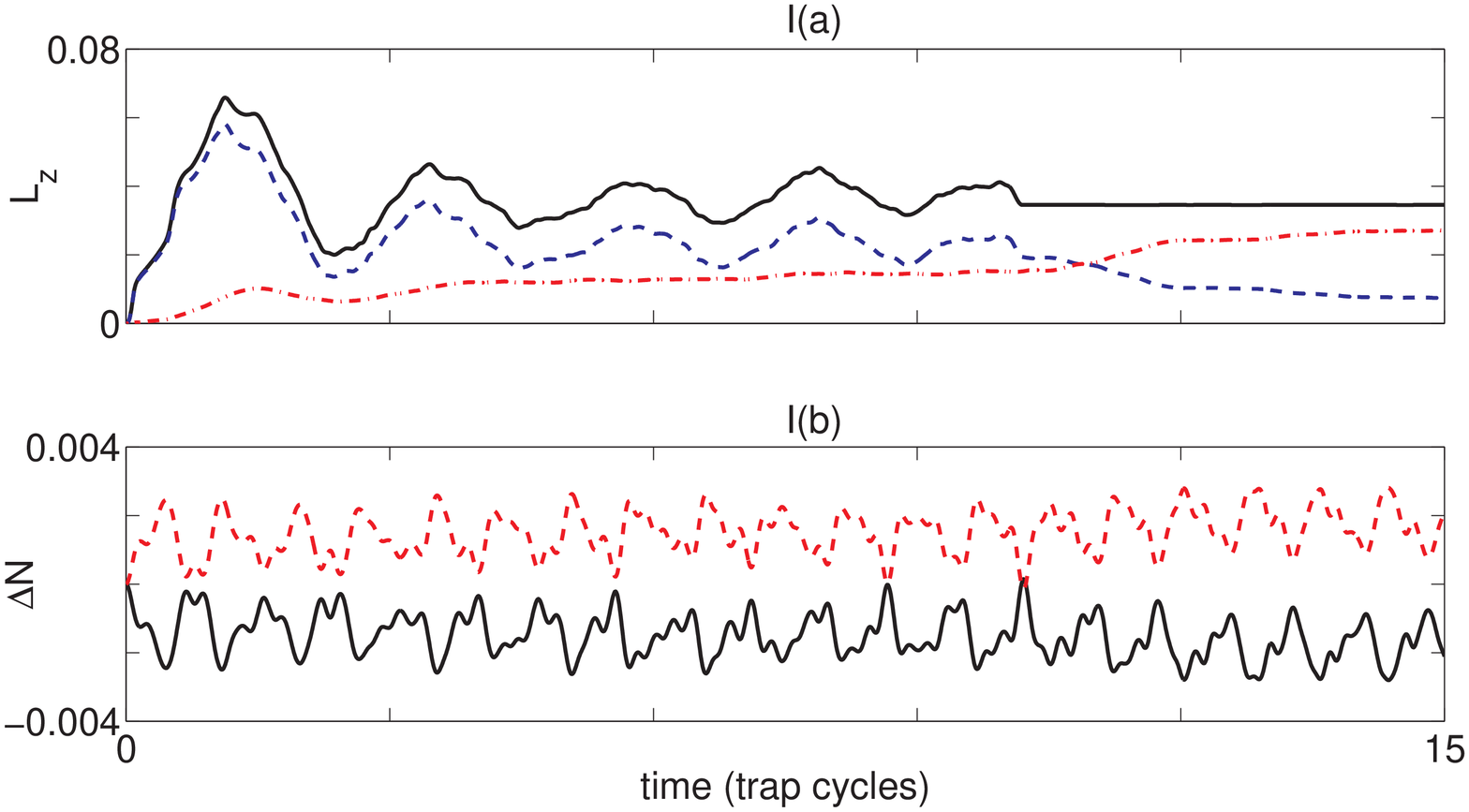}\\
\hspace {.28cm}\includegraphics[scale=0.3]{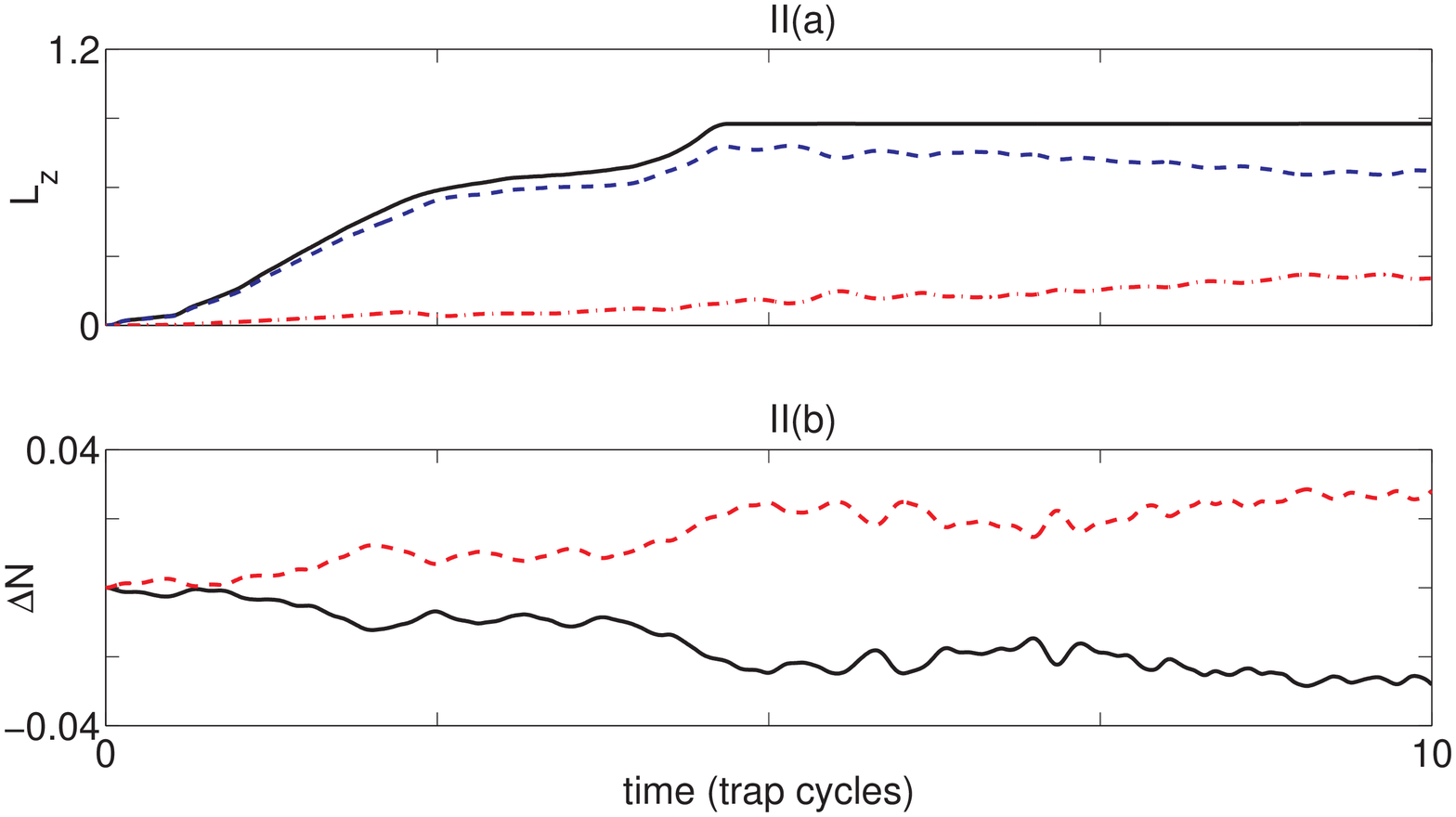}\\
\caption{(Color online) Stirring of condensate (I) using one Gaussian stirrer
at $a$= $\textrm{1.5}r_{0}$ off axis, of amplitude 10 $\hbar\omega_{r}/2$,
FWHM 0.82 $r_{0}$, switched on adiabatically over 0.2 trap cycles,
stirring frequency 0.38 $\omega_{r}$, stirrer on for 10 trap cycles,
and (II) using one Gaussian stirrer at $a$= $\textrm{1.5}r_{0}$
off axis, of amplitude 10 $\hbar\omega_{r}/2$, FWHM 0.82 $r_{0}$,
switched on adiabatically over 0.2 trap cycles, stirring frequency
0.5 $\omega_{r}$, stirrer on for 4.5 trap cycles showing (a) $z$
component of total angular momentum (solid line), condensate angular
momentum (dashed line), and thermal population angular momentum (dash-dotted
line) versus time, (b) change in thermal population and condensate
population (solid line) versus time. The change in the populations
$\Delta N$ is given in units of the number of atoms $N$. All angular
momenta are in units of $N\hbar$.}

\end{figure}
Figure 6 I(a) shows the condensate, thermal, and total angular momentum
in the first case where the stirring frequency ($\Omega=0.38\omega_{r}$)
is sub-critical, and Fig. 6II(a) in the second case where $\Omega=0.5\omega_{r}$
which is above the critical stirring frequency, and where two vortices
are produced as shown in Figs. 4 and 5. In the case of the first simulation,
the rotational frequency $\Omega$ is sub-critical, so no vortices
are produced within the Thomas Fermi radius, therefore negligible
angular momentum is transferred to the condensate. This angular momentum
is due to circulation corresponding to vortices in regions of neglible
density. We see in both cases that a small amount of angular momentum
is also transferred to the thermal cloud. We note that $L_{z}$ remains
fixed once the stirring ceases after 10.2 trap cycles in the first
case, and 4.7 trap cycles in the second in accordance with the conservation
of angular momentum. We note in the first case where the stirring
frequency is subcritical that the angular momentum for the thermal
population increases for the first few trap cycles, reaching a saturation
value after $\sim4\rightarrow5$ trap cycles. Simulations over a much
longer time scale (not presented here) reveal a slow cyclical change
in the thermal population angular momentum of a few percentage points.
Once the stirrer is switched off, we see a steady increase in the
thermal population angular momentum with a resultant loss in the angular
momentum of the condensate (since the overall angular momentum is
conserved). During the time of stirring, the condensate angular momentum
varies cyclically corresponding to the variation of the distance of
vortices from the axis (see also Ref. \cite{key-13e}). In the case
where the stirring frequency is subcritical, these vortices would
lie outside the Thomas-Fermi radius, and would therefore be subthreshold.
Figs. 6I(b) and 6II(b) show, respectively, the changes in condensate
and thermal populations, where we see rigorous observance of particle
number conservation. In order to understand why there exists a critical
frequency for the production of vortices within the Thomas-Fermi radius,
we calculate the Landau critical velocity $v_{LC}$ given by the local
speed of sound at the point of the stirrer $s=\sqrt{ng/m}$, where
$n$ is the density of the condensate and $m$ is the mass of a particle.
We use the local density approximation  and calculate the speed of
sound $s$ at the stirrer using the local density $n$ of the superfluid
at the point of the stirrer. In these simulations, we measure the
two-dimensional density $n^{(2D)}$ where we have integrated out the
axial component, assuming the axial confinement is sufficiently tight
that all axial modes except the lowest can be ignored, and we can
write the local Landau critical frequency in terms of the local two-dimensional
density $n^{(2D)}$ as $V_{LC}=\sqrt{2\sqrt{2}C_{2D}n^{(2D)}}$. The
speed of the stirrer is given by $v_{\mathrm{stir}}=2\pi r_{s}\Omega_{s}$
for a stirrer at radius $r_{s}$ rotating at angular frequency $\Omega_{s}$.
We can rewrite this in terms of the Landau critical velocity as\begin{equation}
v_{\mathrm{stir}}=\frac{2\pi r_{s}\Omega_{s}}{\sqrt{2\sqrt{2}C_{2D}n^{(2D)}}}v_{LC}.\label{eq:49}\end{equation}
Table 1 shows numerically calculated values of the time-averaged $z$-component
quantum expectation value of the angular momentum $\bar{L_{z}}$ and
how this is related to the local stirring velocity in units of the
Landau critical velocity for the stirring angular frequencies 0.39,
0.4, 0.45, and $\textrm{0.5 }\omega_{r}$.%
\begin{table}
\begin{tabular}{|l|l|l|}
\hline 
$\Omega(\omega_{r})$ & $v_{\mathrm{stir}}/v_{LC}$ & $\bar{L_{z}}(N\hbar)$\tabularnewline
\hline
\hline 
0.38 & 1.12 & $\begin{array}{c}
\textrm{0.035}\end{array}$\tabularnewline
\hline 
0.39 & 1.15 & $\begin{array}{c}
\textrm{0.045}\end{array}$\tabularnewline
\hline 
0.4 & 1.18 & \multicolumn{1}{l|}{$\begin{array}{c}
\textrm{0.5 (varies between}\sim\textrm{0.25 and}\sim\textrm{0.7)}\end{array}$}\tabularnewline
\hline 
0.45 & 1.33 & $\begin{array}{c}
\textrm{0.55 (varies between}\sim\textrm{0.3 and}\sim\textrm{0.8)}\end{array}$\tabularnewline
\hline 
0.5 & 1.48 & $\begin{array}{c}
\textrm{0.6 for one vortex,}\sim\textrm{\textrm{0.8 }for two vortices}\\
\textrm{and}\sim\textrm{\textrm{2.2 }for three vortices}\end{array}$ \tabularnewline
\hline
\end{tabular}\caption{Relationship of the local stirrer velocity with the average z-component
of the angular momentum $\bar{L_{z}}$ for time-dependent HFB simulations.
The Landau critical velocity is that corresponding to the density
at the stirrer location, and $n^{(2D)}=0.02\textrm{ }r_{0}^{-2}$.}

\end{table}
We see evidence of a critical stirring angular velocity between 0.39
and 0.4$\omega_{r}$ corresponding to a local stirrer velocity just
slightly above the Landau critical velocity. These results are consistent
with simple GPE simulations (results not presented here).

No experiments have been performed on stirring a BEC using a narrow
rotating probe as considered here but experiments \cite{key-13} have
been performed that show evidence of a critical stirring angular velocity
using elliptical deformation of the trap. Experiments to test the
superfluid critical velocity using the linear motion of a narrow probe
laser beam have also been performed \cite{key-13a,key-13b,key-13c},
finding critical velocities for the onset of dissipation $\sim5-10$
times smaller than the critical velocity inferred from the bulk Bogoliubov
speed of sound. Time-dependent GPE simulations \cite{key-13d} also
indicate that dissipation first occurs at speeds significantly less
than the speed of sound. However, the linear motion of the probe in
these works excite vortex dipole or solitons, and are not therefore
directly applicable to the present work.

In a previous study \cite{key-13e}, similar stirring simulations
were carried out using time-dependent GPE simulations. However, the
critical stirring frequency referred to in these results are the thermodynamic
critical frequency (at which the vortex-free and central vortex states
have equal energies), whereas the results presented here refer to
the critical frequency at which vortices are created within the Thomas-Fermi
radius. Nevertheless, the results of Ref. \cite{key-13e} are consistent
with the simulations performed in this work.

The precessional frequencies for single off-axis vortices were obtained
for simulations deploying the single stirrer at $\Omega=0.5\omega_{r}$
for 2.5 trap cycles. Least squares analysis of the vortex trajectories
as described above reveals that the vortex in the second simulation
developed at a distance of $\sim\textrm{1.7}$ harmonic oscillator
units from the axis, with a precessional frequency of $\sim\textrm{0.42}\omega_{r}$,
in very good agreement with the off-axis time-independent calculations
(see Figs. 1(c) and 1(d) in Ref. \cite{key-12}).

In order to test the predicted precessional frequencies for triangular
vortex arrays, three equi-spaced Gaussian stirrers at $a$= $\textrm{1.5 }r_{0}$
off axis, of amplitude 10 $\hbar\omega_{r}/2$ (i.e. of the same order
as the chemical potential, $\mu\gtrsim\textrm{10}\hbar\omega_{r}/2$),
FWHM 0.82 $r_{0}$, stirring frequency ramped up from zero to $\textrm{0.5}\omega_{r}$
over 0.2 trap cycles, are used to stir the BEC for 2.5 trap cycles.
Thus a symmetrical triangular vortex array was produced. The least-squares
estimates of the precessional frequency of  $\Omega_{LS}=0.7\omega_{r}$,
are in very good agreement with the interpolated values from Figs.
1(a) and 1(b) of  $\Omega\sim\textrm{0.7 }\omega_{r}$.

The above results for the stirring simulations indicate very good
agreement with the values predicted using the continuity equation
for the condensate density  in the time-independent calculations.
These results were attained in the case where the stirrers were introduced
adiabatically. It should be noted that the reason for using three
Gaussian stirrers, as opposed to just one, is not that we need three
stirrers to create three vortices (one stirrer is sufficient for this
purpose provided the BEC is stirred long enough at a sufficiently
high frequency), but rather in order to create a symmetrical triangular
vortex array so the time-independent predictions can be tested.

\section{Breaking of the Axial Symmetry of the Trapping Potential}

\begin{figure}
\includegraphics[scale=0.3]{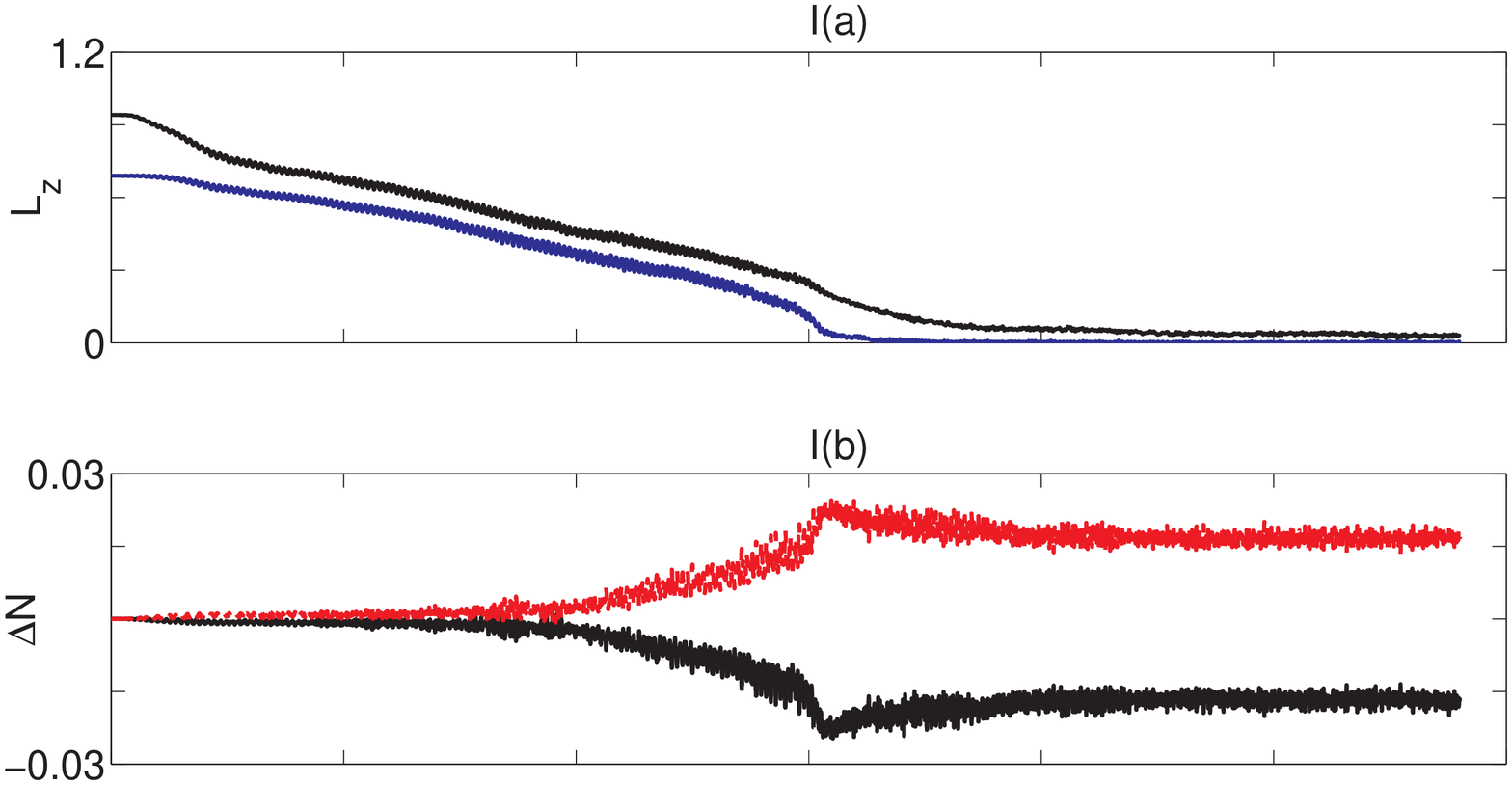}\\
\hspace {.0cm}\includegraphics[scale=0.3]{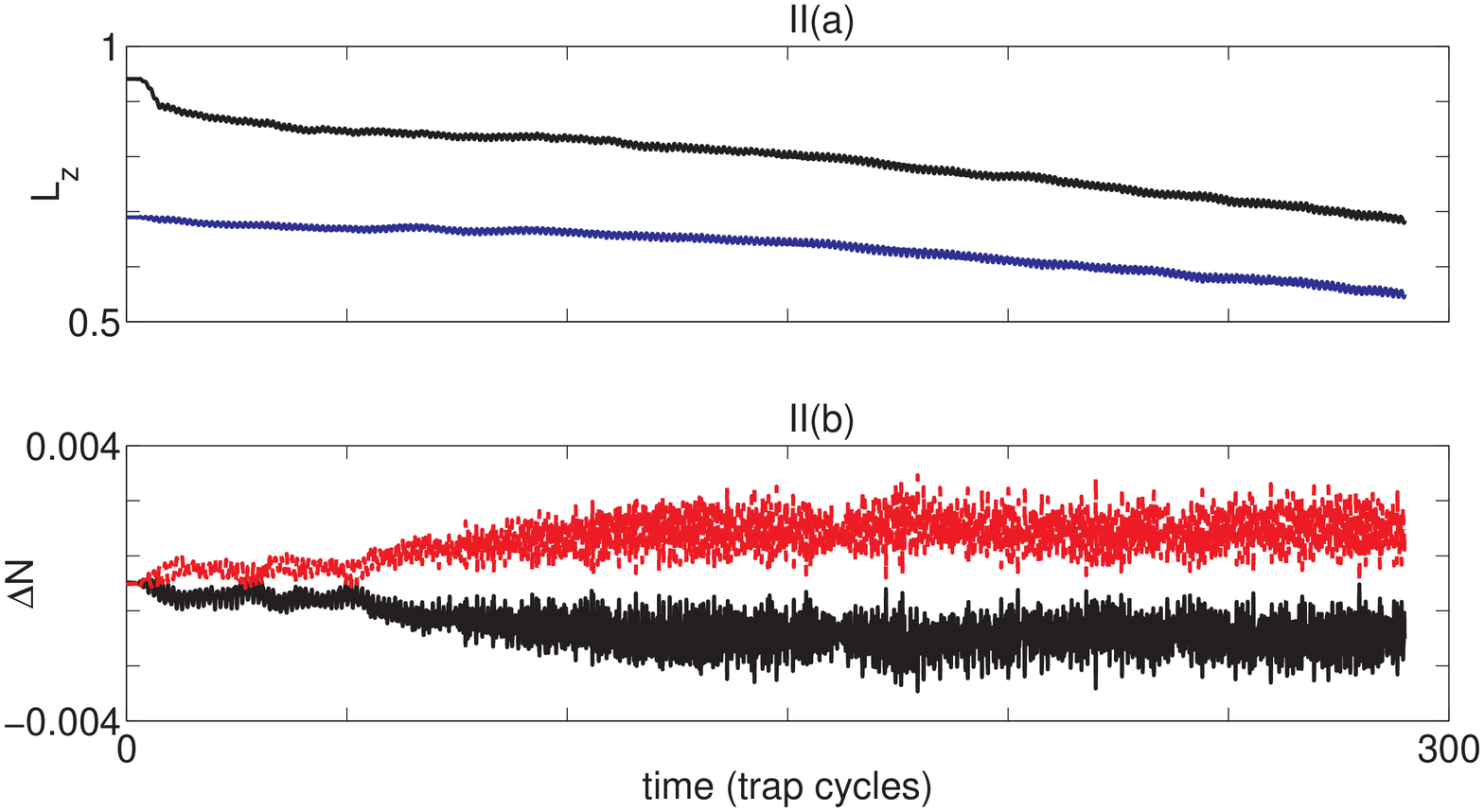}\\
\caption{(Color online) Breaking of axial symmetry; (I) for single off-axis
vortex for turn-on time of 20 trap cycles for eccentricity parameter
$\epsilon$= 0.25, and (II) for single off-axis vortex for turn-on
time of 20 trap cycles for eccentricity parameter $\epsilon$= 0.1
showing (a) $z$-component of total angular momentum and condensate
angular momentum (lower line) versus time and  (b) change in thermal
population and condensate population (lower line) versus time. The
change in the populations $\Delta N$ is given in units of the number
of atoms $N$. All angular momenta are in units of $N\hbar$.}

\end{figure}
\begin{figure}
\includegraphics[scale=0.34]{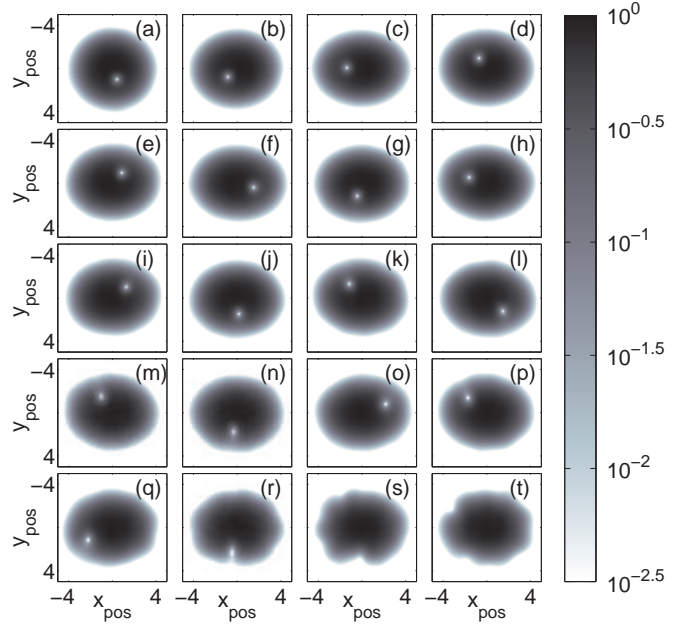}\caption{Simulated absorption images \cite{key-12e} for breaking of axial
symmetry with turn-on time of 20 cycles for single off-axis vortex
for eccentricity parameter $\epsilon$= 0.25 after (a) 8, (b) 16,
(c) 24, (d) 32, (e) 40, (f) 48, (g) 56, (h) 64, (i) 72, (j) 80, (k)
88, (l) 96, (m) 104, (n) 112, (o) 120, (p) 128, (q) 136, (r) 144,
(s) 152, and (t) 160 trap cycles. All positions are in units of the
harmonic oscillator length $r_{0}$.}

\end{figure}
\begin{figure}
\includegraphics[scale=0.32]{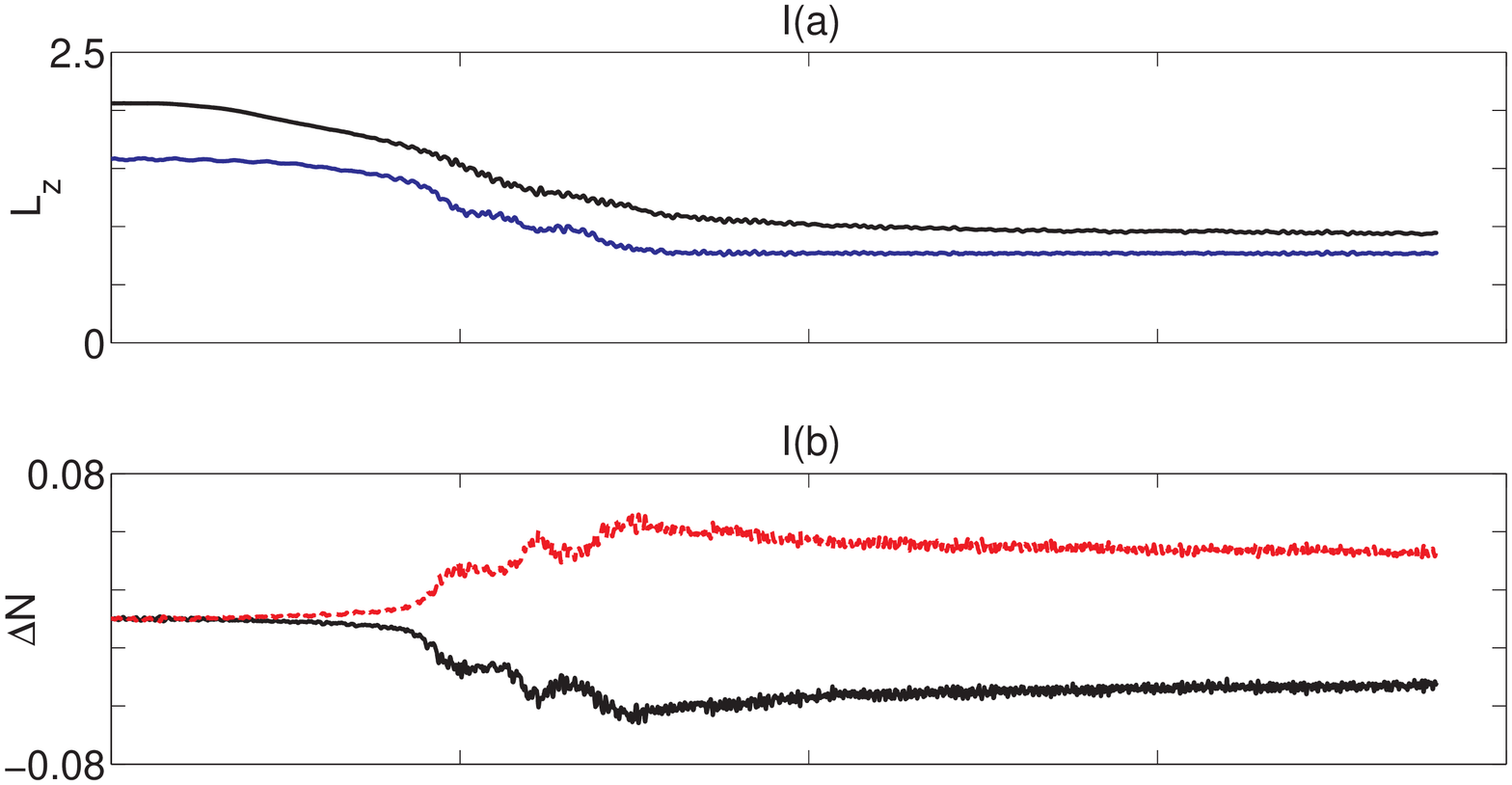}\\
\hspace {.0cm}\includegraphics[scale=0.3]{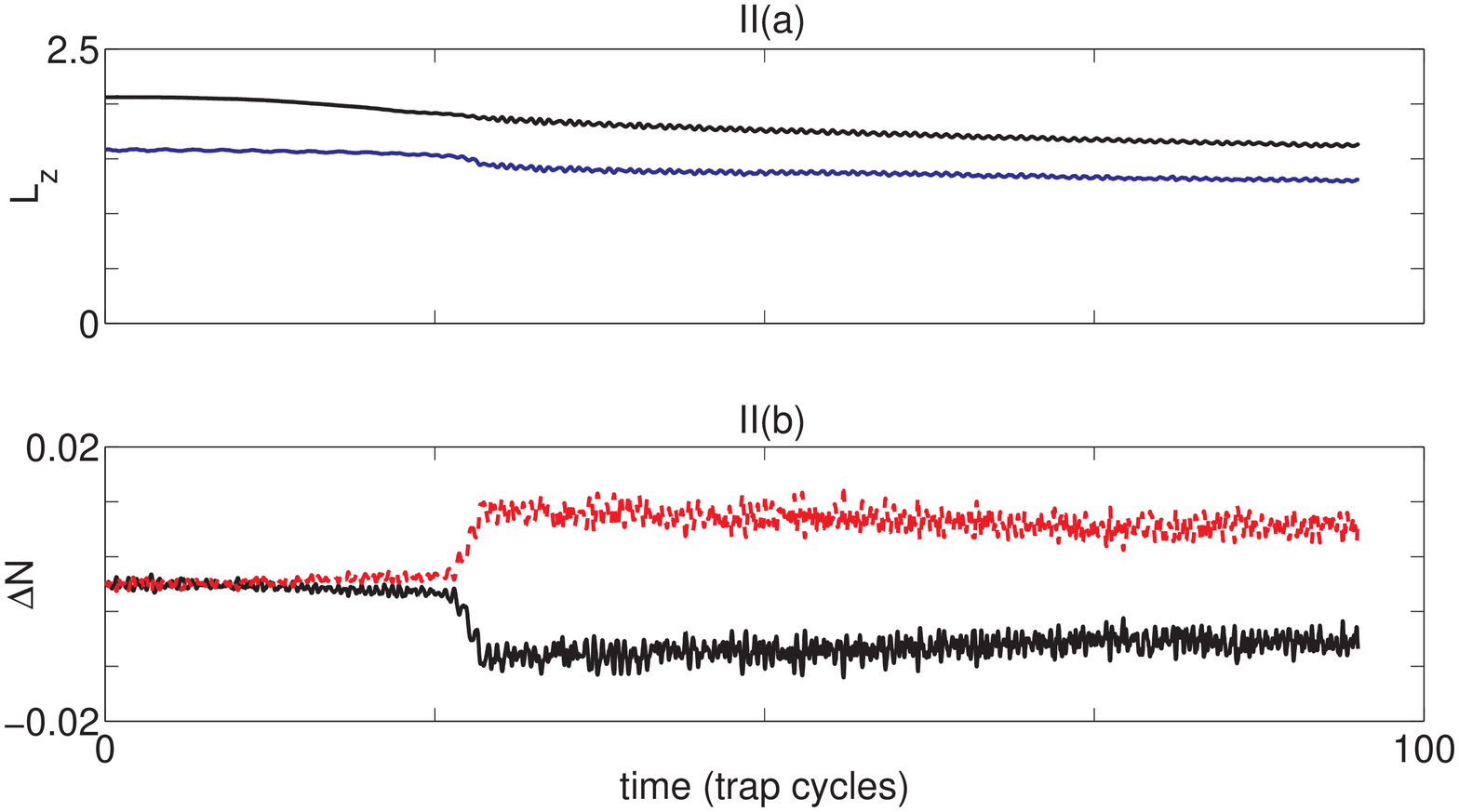}\caption{(Color online) Breaking of axial symmetry, (I) for triangular vortex
array for turn-on time of 20 trap cycles for eccentricity parameter
$\epsilon$= 0.25, and (II) for triangular vortex array for turn-on
time of 20 trap cycles for eccentricity parameter $\epsilon$= 0.1
showing (a) $z$-component of total angular momentum and condensate
angular momentum (lower line) versus time and  (b) change in thermal
population and change in condensate population (lower line) versus
time. The change in the populations $\Delta N$ is given in units
of the number of atoms $N$. All angular momenta are in units of $N\hbar$.}

\end{figure}
\begin{figure}
\includegraphics[scale=0.32]{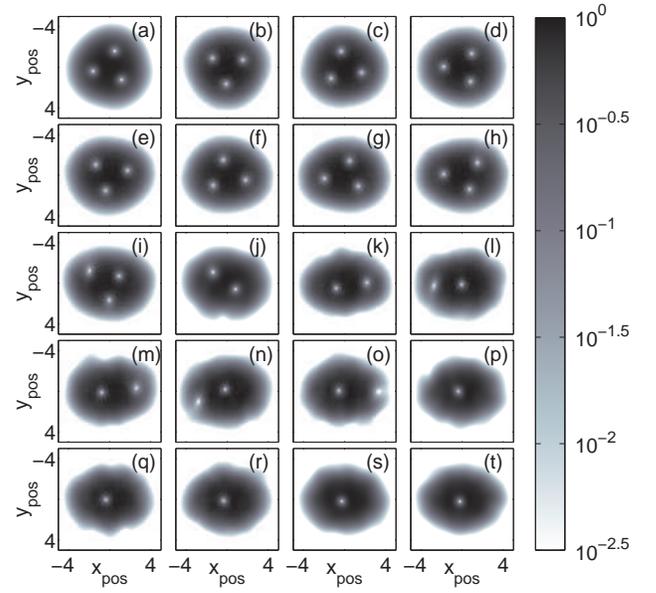}\caption{Simulated absorption images \cite{key-12e} for breaking of axial
symmetry for triangular vortex array over 20 trap periods with lattice
parameter $a$ = 1.65$\sqrt{3}$ for eccentricity parameter $\epsilon$=
0.25 after (a) 2.5, (b) 5, (c) 7.5, (d) 10, (e) 12.5, (f) 15, (g)
17.5, (h) 20, (i) 22.5, (j) 25, (k) 27.5, (l) 30, (m) 32.5, (n) 35,
(o) 37.5, (p) 40, (q) 42.5, (r) 45, (s) 47.5, and (t) 50 trap cycles.
All positions are in units of the harmonic oscillator length $r_{0}$.}

\end{figure}
\begin{figure}
\includegraphics[scale=0.35]{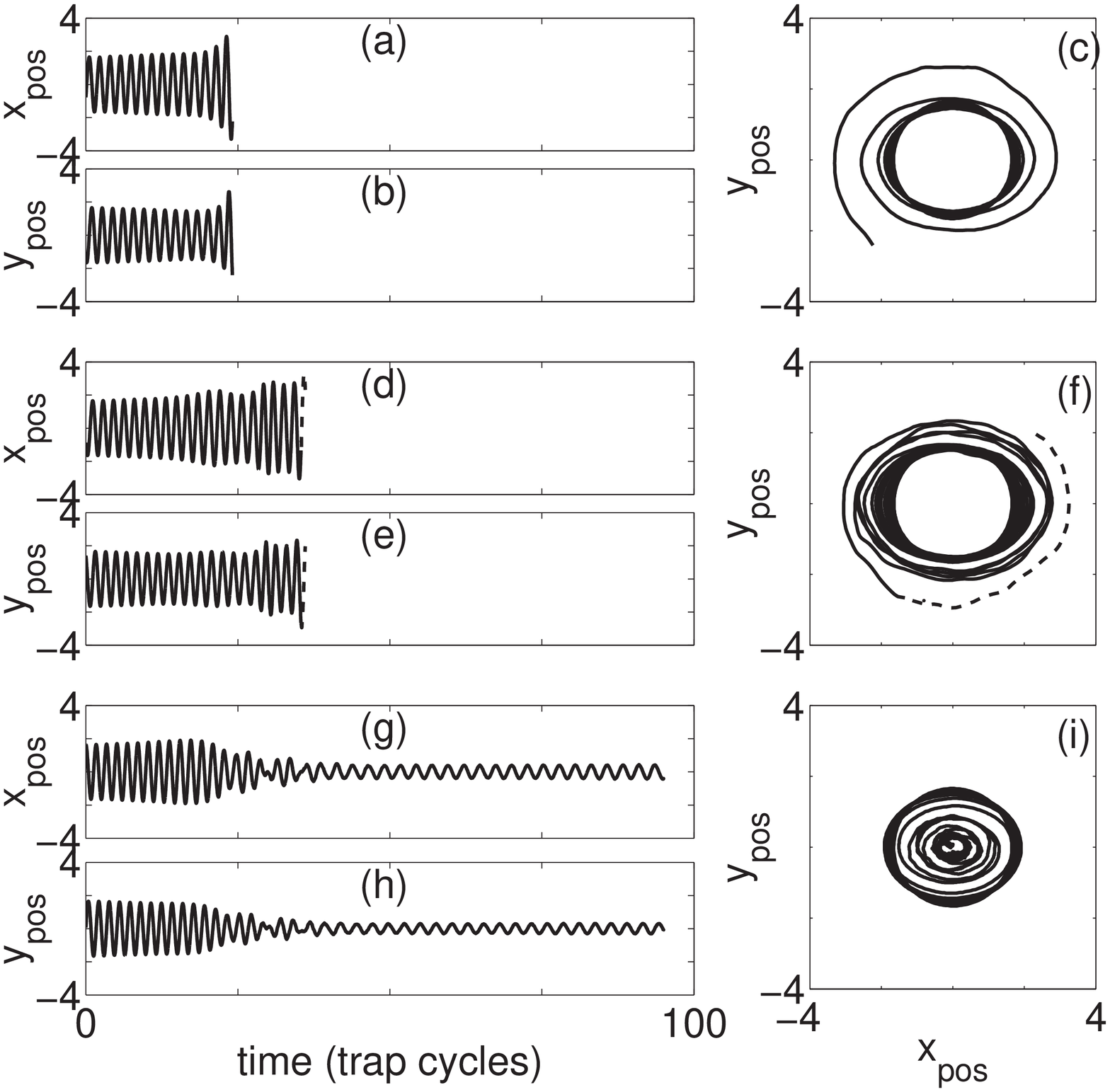}\caption{Breaking of axial symmetry for triangular vortex array for turn-on
time of 20 trap cycles for eccentricity parameter $\epsilon$ = 0.25
showing (a) $x$ displacement of vortex 1, (b) $y$ displacement of
vortex 1, (c) trajectory of vortex 1, (d) $x$ displacement of vortex
2, (e) $y$ displacement of vortex 2, (f) trajectory of vortex 2,
and (g) $x$ displacement of vortex 3, (h) $y$ displacement of vortex
3, and (i) trajectory of vortex 3. All positions are in units of the
harmonic oscillator length $r_{0}$.}

\end{figure}
Noether's theorem \cite{key-15} states that in a conservative physical
system, a differentiable symmetry of the Lagrangian (and, it can be
shown, of the Hamiltonian) has a conserved quantity associated with
the system (i.e a corresponding conservation law). In this case the
axial symmetry of the trapping potential implies the conservation
of the axial component (i.e., the $z$-component) of the angular momentum.
The quantum expectation value of the angular momentum is given by
$\left\langle \mathbf{L}\right\rangle =\int d\mathbf{r}\left\langle \hat{\psi}^{\dag}\hat{\mathbf{L}}\hat{\psi}\right\rangle $,
where $\hat{\psi}=\hat{\Phi}+\hat{\eta}$ as above. Thus using the
expression (\ref{eq:22}) for the non-condensate density matrix, we
find that\begin{equation}
\begin{array}{ccl}
\left\langle \mathbf{L}\right\rangle  & = & -i\hbar\int d\mathbf{r}\left[\Phi^{*}\left(\mathbf{r}\times\mathbf{\nabla}\right)\Phi+\sum_{q}\left(n_{q}u_{q}^{*}\left(\mathbf{r}\times\mathbf{\nabla}\right)u_{q}\right.\right.\\
 &  & \left.\left.+\left(n_{q}+1\right)v_{q}\left(\mathbf{r}\times\mathbf{\nabla}\right)v_{q}^{*}\right)\right].\end{array}\label{eq:52}\end{equation}
It can be shown from the time-dependent HFB equations (\ref{eq:16})
and (\ref{eq:17}) that\begin{equation}
\frac{d\left\langle \mathbf{L}\right\rangle }{dt}=\int d\mathbf{r}\left(\left|\Phi\right|^{2}+\tilde{n}\right)\hat{\mathbf{L}}V_{T}.\label{eq:50}\end{equation}
This determines the rate of change in angular momentum in the BEC
system, and we see that $\frac{d}{dt}L_{z}=0$ for an axially symmetric
trapping potential in accordance with Noether's theorem. The implication
of this is that when this symmetry is broken, $L_{z}$ is no longer
conserved, and since there is no external source of angular momentum
here, we conclude that angular momentum will be lost since $\frac{d}{dt}L_{z}$
will now be non-zero by equation (\ref{eq:50}). This leads to vortex
decay, and the vortex spirals outward as angular momentum is lost.
This is in accordance with work done by Zhuravlev \emph{et. al.} \cite{key-15a}
on the arrest of vortex arrays by a trap anisotropy, although this
work deals with large vortex arrays, whereas the the calculations
performed here concern single off-axis vortices and small vortex arrays.
In order to test this, we introduce an eccentricity into the trapping
potential, thereby breaking the axial symmetry. Initially the trapping
potential is axially symmetric and the eccentricity introduced by
means of the perturbation potential\begin{equation}
V_{\mathrm{per}t}(r,\theta)=\epsilon r^{2}\left(\sin^{2}\theta-\cos^{2}\theta\right)\label{eq:51}\end{equation}
which is introduced adiabatically after two trap cycles over a period
of 20 trap cycles. In these simulations, we introduce eccentricities
of $\epsilon\textrm{ = 0.25}$ and $\epsilon\textrm{ = 0.1}$ in both
the cases of a single off-axis vortex  and of a triangular vortex
array. In the single vortex case, the initial state consists of a
BEC with a single precessing vortex situated at position $a=\textrm{ 1.1}$
trap units from the axis. The $z$-component of the angular momentum
$L_{z}$ versus time in trap cycles is plotted in Figs. 7I(a) and
7II(a) for the cases of $\epsilon\textrm{ = 0.25}$ and $\epsilon\textrm{ =  0.1},$
respectively, where we see a decrease in $L_{z}$ with time once the
symmetry of the trapping potential has been broken. Figures 7I(b)
and 7II(b) show the respective changes in the thermal and condensate
populations, and reveal an increase in the thermal population as the
vortex spirals outwards and finally decays. This is consistent with
the notion that the vortex decays into excitations when it reaches
the condensate boundary \cite{key-15b}. Simulated absorption images
for $\epsilon\textrm{ = 0.25}$ with turn-on time of 20 trap cycles
are shown in Figs. 8(a)-8(t). In the simulations with $\epsilon\textrm{ = 0.25}$
the lifetime of the vortex is approximately 145 to 150 trap cycles,
and for $\epsilon=.1$ substantially larger, with an extrapolated
value of the order of 700 trap cycles. The vortex precessional frequencies
are in good agreement with the time-independent calculations, and
we note that in all cases from the vortex trajectories that the vortices
precess faster as they move outward \cite{key-15c}, which, again,
is in agreement with the time-independent results (see Fig. 1(c) of
Ref. \cite{key-12}), before finally decaying. This is also consistent
with the reduction in angular momentum {[}see Figs. 7I(a) and 7II(a)
for the cases of $\epsilon=0.25$ and $\epsilon=0.1$ respectively].
The vortex moves outwards and precesses faster as it does so in agreement
with Fig. 1(c) of Ref. \cite{key-12}, accompanied by a loss in angular
momentum {[}see Figs. 7I(a) and 7II(a)]. As discussed above, this
loss of angular momentum is due to the symmetry breaking of the trap
 and the loss of the angular momentum is as given by (\ref{eq:50}).
The loss of the vortex after $\sim145\rightarrow150$ trap cycles
(loss of tracking after $\sim146$ trap cycles) for the case $\epsilon=0.25$
is consistent with the decline in $L_{z}$ seen in Fig. 7 I(a) and
the sudden increase in the thermal population in Fig. 7 I(b) at $\sim150$
trap cycles. These results are also in qualitative agreement with
recent simulations using a classical field treatment comprising the
projected Gross-Pitaevskii equation \cite{key-16}.

Similar simulations were performed for the triangular vortex array
with lattice parameter $a=1.65\sqrt{3}$, where the trapping potential
is initially axially symmetric, and the eccentricity is ramped up
to $\epsilon\textrm{ = .25}$ and $\epsilon\textrm{ = .1}$ over a
period of 20 trap cycles.

The angular momentum calculated using Eq. (\ref{eq:52}) is plotted
in Fig. 9 I(a) for the case of $\epsilon\textrm{ = 0.25}$ and in
Fig. 9 II(a) for the case of $\epsilon\textrm{ = 0.1}$. Figures 9I(b)
, and 9II(b) show the respective changes in the thermal and condensate
populations and reveal increases in thermal population as each successive
vortex spirals outward and finally decays, again in accordance with
decay of the vortex into excitations when it reaches the condensate
boundary \cite{key-15b}. The simulated absorption images for $\epsilon=.25$
are shown in Figs. 10(a)-10(t), and the vortex trajectories in Figs.
11(a)-11(i). We note from the trajectory of vortex 3, shown in Figs.
11(g) and 11(h), the decrease in the precessional frequency following
the departure of a vortex, in agreement with the time-independent
predictions for vortex arrays shown in Figs. 3(a) and 3(b), which
show the dependence of the precessional frequencies of the vortex
array on the lattice parameter for the cases of three vortices (our
initial configuration), and of two vortices, and also with Figs. 1(a)
and (b) which give the time-independent predicted precessional frequencies
for the triangular vortex array at various temperatures and values
of lattice parameter. The prediction here is $\Omega\sim.6\omega_{r}$
for lattice parameter $a\sim2\sqrt{3}=3.46$ and is in good agreement
with the measured time-dependent value {[}see Figs. 11(a) and 11(b)
for $x_{pos}$ versus $T$ and $y_{pos}$ versus $T$ and the time-independent
predictions in Figs. 1(a) and (b)]. We also see good agreement with
the measured value of the precessional frequencies of a single vortex
(after the other two vortices have left the condensate) indicated
in Figs. 11(g) and 11(h), and the time-independent predictions given
in Figs. 1(c) and 1(d) in Ref. \cite{key-12}. We note that Figs.
10 and 11, which show the vortex decay in the case $\epsilon=.25$,
are consistent indicating a decay time of $\sim24$ trap cycles for
the first vortex and a decay time of $\sim38$ trap cycles for the
second. This is also consistent with Fig. 9I which shows sudden changes
in the thermal population at these times, and a steady decrease in
$L_{z}$ with slight drops accompanying the decay of each vortex.
We also note a sudden change in the thermal population after $\sim30$
trap cycles. This accompanies the complex precessional motion of the
third vortex at $\sim30$ trap cycles as shown in Figs. 11(g) and
11(h) due to the interaction between the vortices. The lifetimes of
the vortices in the case of the triangular vortex for $\epsilon=.25$
vary considerably - the first vortex to leave the condensate has a
lifetime of between $\sim\textrm{20}$ and $\sim\textrm{25}$ trap
cycles, while the second has a lifetime varying between $\sim\textrm{25}$
and $\sim\textrm{38}$ depending on the scenario. In all cases the
loss of a vortex is characterized by a slight drop in $L_{z}$ and
sudden increases in the thermal population with a corresponding drop
in the condensate population (see Fig. 9). The departure of the vortices
is also marked by the trajectories of the vortices shown in Figs.
11(a)-11(f) where we see the end of the vortex tracks. The lifetime
of the remaining vortex can be inferred from Fig. 9 I(a) only by extrapolation
and is of the order of 300-400 trap cycles {[}since the residual angular
momentum (i.e. the angular momentum at the time that only a single
vortex remains) is higher than in the single vortex case]. In the
case of the triangular vortex array for $\epsilon\textrm{ = 0.1}$,
the lifetime of the first vortex is of the same order ($\sim25\rightarrow30$
trap cycles), but the lifetimes for the second and third vortices
are considerably longer, with the second vortex having an extrapolated
lifetime of $\sim\textrm{400}$ trap cycles {[}see Fig. 9 II(a)].
It is impossible to ascertain the lifetime of the remaining vortex
from the simulations, but it is probably of the order of a few thousand
trap cycles that is impractical to simulate with the facilities available.

\section{Summary}

We have developed a novel method utilizing the continuity equation
for the condensate density  to make predictions of the precessional
frequency of single off-axis vortices, and of vortex arrays in Bose-Einstein
condensates (BECs) at finite temperature. We also presented an orthogonalized
Hartree-Fock-Bogoliubov (HFB) formalism for which a zero-energy eigenvalue
exists, in contrast to the standard HFB theory. We solved the continuity
equation for the condensate density and the time-independent orthogonalized
HFB equations self-consistently in the frame rotating at the precessional
frequency. Thus we were able to find stationary solutions for quasi-two-dimensional
rotating systems in the corotating frame. We compared these results
with time-dependent predictions where we simulated stirring of the
condensate, finding good agreement with the predicted precessional
frequencies. These results are completely consistent with GPE simulations,
and are in qualitative agreement with the projected Gross-Pitaevskii
simulations and with the areal density law. We also verified that
angular momentum is conserved in the quasi-two-dimensional system
for an axially symmetric trapping potential and that breaking this
symmetry leads to the loss of angular momentum and hence, to the decay
of vortices.\\

This work was funded through the New Economy Research Fund contract
NERF-UOOX0703: Quantum Technologies.


\begin{thebibliography}{10}
\bibitem{key-1}A. L. Fetter, Rev. Mod. Phys., \textbf{81}, 647 (2009).

\bibitem{key-2}D. A. W. Hutchinson, K. Burnett, R. J. Dodd, S. A.
Morgan, M. Rusch, E. Zaremba, N. P. Proukakis, M. Edwards, and C.
W. Clark, J. Phys. B: At. Mol. Opt. Phys. \textbf{33}, 3825 (2000).

\bibitem{key-3}D. A. W. Hutchinson, E. Zaremba, and A. Griffin, Phys.
Rev. Lett. \textbf{78}, 1842 (1997).

\bibitem{key-4}B. G. Wild, P. B. Blakie, and D. A. W. Hutchinson,
Phys. Rev. A \textbf{73}, 023604 (2006).

\bibitem{key-5}A. Griffin, Phys. Rev. B \textbf{53}, 9341 (1996).

\bibitem{key-6}I. Coddington, P. C. Haljan, P. Engels, V. Schweikhard,
S. Tung, and E. A. Cornell, Phys. Rev. A \textbf{70}, 063607 (2004).

\bibitem{key-6a}D. E. Sheehy and L. Radzihovsky, Phys. Rev. A \textbf{70},
051602(R) (2004).

\bibitem{key-7}S. Tung, V. Schweikhard, and E. A. Cornell, Phys.
Rev. Lett. \textbf{97}, 240402 (2006).

\bibitem{key-8}H. Buljan, M. Segev and A. Vardi, Phys. Rev. Lett.
\textbf{95}, 180401 (2005) ; T. Ernst and J. Brand, Phys. Rev. A \textbf{81},
033614 (2010).

\bibitem{key-9}S. Wuster, J. J. Hope, and C. M. Savage, Phys. Rev.
A \textbf{71}, 033604 (2005) ; S. Wuster, B. J. Dkabrowska-W$\ddot{\textrm{u}}$ster,
S. M. Scott, J. D. Close and C. M. Savage, Phys. Rev. A \textbf{77},
023619 (2008) ; Milstein J. N., Menotti C., and Holland M. J., New
J. Phys. 5, 52 (2003).

\bibitem{key-10}Y. Castin and R. Dum, Rhys. Rev. A \textbf{57}, 3008
(1998).

\bibitem{key-10a}This result applies for any linear differential
operator $\hat{\mathbf{L}}$, where $\hat{\mathbf{L}}=i\hbar\left(\mathbf{r}\times\mathbf{\nabla}\right)$
for angular momentum and $\hat{\mathbf{L}}=i\hbar\nabla$ for linear
momentum.

\bibitem{key-11}N. M. Hugenholtz and D. Pine, Phys. Rev. \textbf{116},
489 (1959).

\bibitem{key-11a}D. S. Petrov, M. Holzmann, and G. V. Shlyapnikov,
Phys. Rev. Lett. \textbf{84}, 2551 (2000).

\bibitem{key-12}B. G. Wild and D. A. W. Hutchinson, Phys. Rev. A
\textbf{80}, 035603 (2009).

\bibitem{key-12a}H. M. Nilsen, G. Baym, and C. J. Pethick, Proc.
Natl. Acad. Sci. USA, \textbf{103}, 7978 (2006).

\bibitem{key-12b}D. M. Jezek and H. M. Cataldo, Phys. Rev. A \textbf{77},
043602 (2008).

\bibitem{key-12c}M. Guilleumas and R. Graham, Phys. Rev. A \textbf{64},
033607 (2001).

\bibitem{key-12d}N. Papanicolaou, S. Komineas, and N. R. Cooper,
Phys. Rev. A \textbf{72}, 053609 (2005).

\bibitem{key-12e}The simulated absorption images shown are logarithmic
plots given by $n_{plot}=\log\left(n/n_{max}\right)$, where $n$
refers to the specific density to be plotted and $n_{max}$ is the
overall maximum density of the BEC (i.e., the maximum of sum of the
condensate and non-condensate densities).

\bibitem{key-12d1}It should be noted that in all cases the precessional
frequency of the vortices nucleated in the stirring is distinct from
the rotational frequency of the stirrer itself - see for example,
Fig. 5 which shows clearly that these precessional frequencies are
uncorrelated with the stirring frequency $\Omega_{stir}=0.5\omega_{r}$.

\bibitem{key-13e}B. M. Caradoc-Davies, R. J. Ballagh and K. Burnett,
Phys. Rev. Lett. \textbf{83}, 895 (1999).

\bibitem{key-13}F. Chevy, K. W. Madison, and J. Dalibard, Phys. Rev.
Lett. \textbf{85}, 2223 (2000).

\bibitem{key-13a}C. Raman, M. K$\ddot{o}$hl, R. Onofrio, D. S. Durfee,
C.E. Kuklewicz, Z. Hadzibabic, and W. Ketterle, Phys. Rev. Lett. \textbf{83},
2502 (1999).

\bibitem{key-13b}R. Onofrio, C. Raman, J. M. Vogels, J. R. Abo-Shaeer,
A. P. Chikkatur, and W. Ketterle, Phys. Rev. Lett. \textbf{85}, 2228
(2000).

\bibitem{key-13c}P. Engels and C. Atherton, Phys. Rev. Lett. \textbf{99},
160405 (2007).

\bibitem{key-13d}B. Jackson, J. F. McCann, and C. S. Adams, Phys.
Rev. A \textbf{61}, 051603 (2000).

\bibitem{key-15}Noether E., Nachr. D. K$\ddot{\textrm{o}}$nig. Gesellensch.
D. Wiss. Zu G$\ddot{\textrm{o}}$ttingen, Math-Phys. Klasse \textbf{2}
235-257 (1918) ; M. A. Peskin and D. V. Schroeder, \emph{An Introduction
to Quantum Field Theory}, (Addison-Wesley, Reading, MA, 1995).

\bibitem{key-15a}O. N. Zhuravlev, A. E. Muryshev, and P. O. Fedichev,
Phys. Rev. A \textbf{64}, 053601 (2001). 

\bibitem{key-15b}P. O. Fedichev and G. V. Shlyapnikov, Phys. Rev.
A \textbf{60}, R1779 (1999). 

\bibitem{key-15c}This is a well-known property of solutions of the
GPE \cite{key-12c,key-12d}, provided one is away from the lowest-Landau-Level
limit. It is interesting to note that this behavior persists at finite
temperature in the HFB solutions.

\bibitem{key-16}T. M. Wright, A. S. Bradley, and R. J. Ballagh, Phys.
Rev. A \textbf{81}, 013610 (2010). 
\end{thebibliography}
\end{document}